\documentclass[prd,aps,a4paper,eqsecnum,floatfix,twocolumn,nofootinbib]{revtex4}  

\usepackage{graphicx} 
\usepackage{amsmath,amsfonts,amssymb}
\usepackage{placeins}
\usepackage{appendix}
\usepackage[caption=false,justification=justified]{subfig}

\newcommand{\beq}{\begin{equation}}
\newcommand{\eeq}{\end{equation}}
\newcommand{\bea}{\begin{eqnarray}}
\newcommand{\eea}{\end{eqnarray}}
\newcommand{\be}{\begin{equation}}      
\newcommand{\ee}{\end{equation}}
\newcommand{\GN}{G_{\text{N}}}

\begin{document}

\title{Scalar self-force effects in neutral $W$-soliton backgrounds}

\author{Massimo Bianchi$^{1}$, Donato Bini$^{2}$, Giorgio Di Russo$^{3}$}
  \affiliation{
$^1$Dipartimento di Fisica, Universit\`a di Roma \lq\lq Tor Vergata" and Sezione INFN Roma2, Via della
Ricerca Scientifica 1, 00133, Roma, Italy\\
$^2$Istituto per le Applicazioni del Calcolo ``M. Picone,'' CNR, I-00185 Rome, Italy\\
$^3$School of Fundamental Physics and Mathematical Sciences, Hangzhou Institute for Advanced Study, UCAS, Hangzhou 310024, China\\
}

\date{\today}

\begin{abstract}
We investigate several geometrical and physical properties of the recently found $W$-soliton solution (neutral case). 
We discuss both the genuine 5d solution and its reduction to 4d and highlight similarities and differences.
In both cases, we study scattering processes of massless and massive particles in  the background, reconstructing the gauge-invariant scattering angle, either with exact expressions or with large-angular momentum expansion expressions, which we show how to resum in a useful form.
Finally, we analyze the propagation of a test scalar field in the $W$-soliton background and compute the spectrum of Quasi Normal Modes in the case of (non-)minimal coupling and the radiated energy in the case of minimal coupling. Our result for the energy loss is fully analytic and presented in a Post-Newtonian  expansion, following the approach termed gravitational self force.
\end{abstract}

\maketitle
\section{Introduction}
\label{Intro}

Finding viable descriptions for the micro-states of (neutral, rotating) Black Holes (BHs) is a very active endeavour \cite{Lunin:2001jy, Bena:2015bea, Bianchi:2018kzy,Bena:2019azk}. Recently significant progress has been made with the discovery of the so-called $W$-soliton \cite{Chakraborty:2025ger, Dima:2025tjz},
an exact solution of 5d Einstein-Maxwell (E-M) equations. The main virtue of these solution, as compared to Topological Stars \cite{Bah:2020ogh, Bah:2020pdz, Heidmann:2022ehn, Bianchi:2022qph, Heidmann:2023ojf,  Bianchi:2023sfs, Bianchi:2024vmi, DiRusso:2024hmd} and their rotating versions \cite{Bianchi:2025uis, Heidmann:2025pbb} is that $W$-soliton exists in mass-charge regimes where BHs exist, too. So they can be taken as valid (charged) non-rotating BH micro-states. Quite remarkably their form is extremely simple and tractable, especially in the neutral case. 

Exact solutions of Einstein Equations (EEs) in 4d or higher dimensions are very few which are also physically relevant. Then, even if one were to consider $W$-solitons simply as mathematical solutions, one could get crucial information about gravity in a strong field regime. 

The neutral 5d $W$-soliton can be reduced  to an exact spherically symmetric solution of 4d EEs, sourced by a single (complex) scalar field.
Due to spherical symmetry, the 4d neutral $W$-soliton turns out to be very simple and, at the same time, rich of interesting geometrical and physical properties which we will try and examine. First of all, it approaches the Schwarzschild BH asymptotically, but it exposes a naked singularity, showing the typical repulsive-gravity effects as soon as one approaches it. Clearly the 4d naked singularities is a symptom of the genuine 5d nature of the solution, whereby it is perfectly smooth and horizonless. 

Apart from some geometrical characterization of the spacetime, that extends the analyses in \cite{Chakraborty:2025ger,Dima:2025tjz}, we study the scattering of (neutral) massless and massive particles off the $W$-soliton. We compute the scattering angle both in its exact form (in terms of elliptic integrals) and in a large-angular momentum expansion limit, at the geodesic level, i.e., without back-reaction which will be studied next, after developing the necessary formalism. We display useful resummations for the expanded forms, in terms of  hypergeometric functions.
The antiderivative with respect to the angular momentum of the scattering angle is the radial action, which is then simply obtained  also in a resummed form.
In turn, the radial action is related to the behaviour at infinity of the Jeffreys-Wentzel-Kramers-Brilluein  (JWKB) solution for a  test scalar field wave: this behaviour (monodromy) is fully accounted by the so called \lq\lq renormalized angular momentum" $\nu$ which can be computed according to different approaches,
 e.g., Mano-Suzuki-Takasugi (MST) \cite{Mano:1996vt} and modern gauge theory approach \cite{Aminov:2020yma,Aminov:2023jve,Bautista:2023sdf,Bonelli:2022ten,Bonelli:2021uvf,Bianchi:2021xpr,Bianchi:2021mft,Consoli:2022eey,Fucito:2023afe,Cipriani:2024ygw,Bianchi:2023rlt, DiRusso:2025qpf}, based on quantum Seiberg-Witten (qSW) curves \cite{Nekrasov:2009rc,Nekrasov:2002qd}, that give complementary results in full agreement with one another. We display the Post-Newtonian expansion of $\nu$ for various values of the angular momentum number $\ell$ and check its identification with $a_{\rm SW}$ up to a shift by 1/2.
We also briefly discuss the Quasi Normal Modes (QNMs) with different overtones in the 4d case, by using different approaches: JWKB,  Leaver, numerical integration. For the specific case of neutral $W$-solitons, our results match well with the more general results of \cite{Chakraborty:2025ger, Dima:2025tjz}.   

The main accomplishment of this work concerns in the study of a scalar field perturbing the background spacetime in 5d within the framework of gravitational self force (GSF). We limit our considerations to the minimal coupling case and  to the massless scalar field case, as a proxy for the more interesting but more involved case of gravitational waves. The evolution equations are separated 1) in their angular dependence by expanding the field in (scalar) spherical harmonics; 2) in time, by Fourier transforming the time variable. One is left then with a single radial equation which is of the Confluent Heun type (CHE).
We develop the analogue of the MST formalism (valid and by now almost standardized for BHs) to this special solution (see \cite{Bianchi:2024vmi} for a similar treatment discussing the case of another special solution in 5d, called the ``Topological Star" or ``Top Star'' for short, see e.g.,  \cite{Bianchi:2023sfs,DiRusso:2024hmd,Cipriani:2024ygw,Bena:2024hoh,Cipriani:2025ikx,Bini:2025qyn}) and then compute the energy  loss when the source is moving along a time-like circular equatorial geodesic \footnote{The angular momentum loss is proportional to the energy loss in this circular orbit case.}.

The rest of the paper is organized in two main parts. In the first part we study the 4d case, that is more than a simple warm-up exercise. Then we pass to consider the 5d case, the really important case. A whole section is  devoted to the computation of the energy loss in scalar waves when the source is a neutral massive scalar particle moving along a circular geodesic. The results are compared with the ones for other backgrounds: the Schwarzschild BH and the Top Star. Finally, we make our concluding remarks and discuss  plans for future work. 

We use mostly plus signature convention and units such $(8\pi) \GN=1=c$, unless differently specified.

\section{4d $W$ neutral soliton}

Let us start by analyzing the properties of the neutral $W$-soliton from the reduced  `bottom-up' 4d perspective. The metric \cite{Chakraborty:2025ger, Dima:2025tjz} reads
\bea
ds^2=-fdt^2+\frac{dr^2}{f}+fr^2(d\theta^2+\sin^2\theta d\phi^2)\,,
\eea
with
\beq
f=\sqrt{1-\frac{4M}{r}}\sim 1-\frac{2M}{r}+O\left(\frac{1}{r^2} \right)\,.
\eeq 
This solution approaches the familiar Schwarzschild BH solution  as $r\to \infty$ (far region), but it does not represent a black hole. 
It is sourced by a complex scalar field
\beq
z= \frac{1}{\left(1-\frac{2 M}{r}\right)}\left(\frac{2 M}{ r}+i f \right)\,,
\eeq
with non-canonical kinetic term such that\footnote{One can have a more elegant form for $T_{\mu\nu}$ by defining $\frac{1}{{\rm Im}\, z}\partial_\mu z=\partial_\mu \zeta$ since both $z$ and $\zeta$ will only depend on the radial variable $r$. However, this will remain only a formal simplification.}
\beq
T_{\mu\nu}=\frac{3}{4({\rm Im}\, z)^2}\left(N_{\mu\nu}-\frac12 ({\rm Tr}N) g_{\mu\nu}\right)=\frac{3}{4({\rm Im}\, z)^2}N_{\mu\nu}^{\rm TR}\,,
\eeq
where we have introduced the combinations 
\bea
N_{\mu\nu}&=&\partial_\mu z^* \partial_\nu z+\partial_\mu z \partial_\nu z^*=2\partial_{(\mu} z^* \partial_{\nu )} z \,,\nonumber\\
{\rm Tr}N&=&g^{\mu\nu}N_{\mu\nu}\,,
\eea
as well as 
\beq
X^{\rm TR}_{\mu\nu}=X_{\mu\nu}-\frac12 ({\rm Tr} X) g_{\mu\nu}
\eeq
for the trace-reversed form of a symmetric 2-tensor $X$.
Explicitly, the non-vanishing components of $T_{\mu\nu}$ are
\beq
\left[T_{00},T_{rr},T_{\theta\theta},T_{\phi\phi}\right]=\frac{3M^2}{r^4f^2}\left[-1,-\frac{1}{f^2},r^2,r^2\sin^2\theta  \right]\,,
\eeq
and the EE are written in units such that
\beq
G_{\mu\nu}=    T_{\mu\nu}\,.
\eeq
The Khretschmann invariant is given by
\beq
{\mathcal K}=R_{\mu\nu\rho\sigma}R^{\mu\nu\rho\sigma}=  \frac{12 M^2 (69 M^2-32 M r+4 r^2)}{ r^8 (1-\frac{4 M}{r})^3}
\eeq
implying that $r=0$ and $r=4M$ are both singularities of the present metric. Actually, the (naked) singularity at $r=4M$ limits the spacetime region we will explore here. 
Recall that the naked singularity  is an artefact of the 4d description. The full 5d geometry is completely smooth and horizonless. At $r=4M$ the radius of the fifth dimension shrinks to zero size.
Rescaling ${\mathcal K}$  by a factor $M^4$ and denoting $u=M/r$ one finds the equivalent expression
\bea
M^4 {\mathcal K}&=&-\frac{417}{4096} u -\frac{171}{512} u^2 -\frac{141}{128} u^3-\frac{237}{64} u^4-\frac{207}{16} u^5 \nonumber\\
&+&\frac{417}{4096} \frac{u}{(1-4u)}-\frac{75}{1024} \frac{u^2}{(1-4u)^2}\nonumber\\
&+& \frac{15}{256} \frac{u^3}{(1-4u)^3}\,,
\eea
showing the \lq\lq different weights" for $u=0$ and $4u=1$.
Finally, the Ricci scalar is given by
\beq
R = \frac{6 M^2}{ f^3 r^4}\,.
\eeq
Static observers, i.e., observers  at rest with respect to the coordinates, are considered our \lq\lq fiducial observers." Their four velocity reads
\beq
\label{U_obs}
U=\frac{1}{\sqrt{f}}\partial_t\,,
\eeq
and they are accelerated radially outward, with acceleration
\beq
\label{acc_U}
a(U)= \frac{M}{r^2 f}\partial_r\,, 
\eeq
while both their expansion and their vorticity vanish identically \cite{Jantzen:1992rg}.
Noticeably $a(U)^r>0$ and in order to stay at a fixed position these observers need some inward acceleration, that counter-balance the repulsive effects induced by the presence of a naked singularity. The latter, however, manifests itself causing the magnitude of the necessary acceleration 
\beq
||a(U)||= \frac{M}{r^2 f^2}
\eeq
to increase more and more as soon as one is approaching $r=4M$.
In addition, these observers play a role in the characterization of the $W$-soliton geometry. In fact, for example
\beq
R_{\mu\nu}=6 a_\mu a_\nu\,,\quad G_{\mu\nu}=6 [a_\mu a_\nu]^{\rm TR}\,.
\eeq
In the present case $a(U)$ is purely radial. It is then convenient to work with frame components with respect to and adapted frame to  
the fiducial observers $U=e_{\hat t}$, Eq. \eqref{U_obs}, 
\bea
\label{spat_frame}
e_{\hat r}&=&\sqrt{f} \partial_r \,,\quad
e_{\hat \theta} = \frac{1}{r\sqrt{f}} \partial_\theta \,,\quad
e_{\hat \phi}= \frac{1}{r\sin \theta \sqrt{f}} \partial_\phi\,.\qquad
\eea
Therefore
\beq
\label{R_and_G}
R_{\mu\nu}=\frac{6 M}{r^2 f^2} [e_{\hat r}\otimes e_{\hat r}]_{\mu\nu}\,,\quad G_{\mu\nu}=\frac{6 M}{r^2 f^2} [e_{\hat r}\otimes e_{\hat r}]_{\mu\nu}^{\rm TR}\,,
\eeq
($G_{\mu\nu}=T_{\mu\nu}$ according to our notation) and the trace-reverse operation leads to  
\bea
[e_{\hat r}\otimes e_{\hat r}]^{\rm TR}&=& e_{\hat r}\otimes e_{\hat r}-\frac12 [-e_{\hat t}\otimes e_{\hat t}+e_{\hat r}\otimes e_{\hat r}\nonumber\\
&+& e_{\hat \theta}\otimes e_{\hat \theta}+e_{\hat \phi}\otimes e_{\hat \phi}]\nonumber\\
&=& -\frac12 [-e_{\hat t}\otimes e_{\hat t}-e_{\hat r}\otimes e_{\hat r}\nonumber\\
&+& e_{\hat \theta}\otimes e_{\hat \theta}+e_{\hat \phi}\otimes e_{\hat \phi}]\,.
\eea
Eqs. \eqref{R_and_G} imply that the non Ricci-flatness of this spacetime only occurs in the radial direction.

Null and massive geodesics will be studied in the next subsection, with special attention to scattering processes.

\subsection{Null geodesics}

Thanks to spherical symmetry we can limit our consideration to the equatorial geodesics  $\theta=\pi/2$, which are simply determined by the null condition $ds^2=0$, i.e., 
\beq
\label{null_met}
-f dt^2 + {dr^2\over f} + f r^2 d\phi^2=0\,.
\eeq
Denoting by a dot the differentiation with respect to an affine parameter $\tau$ we find
the two Killing relations
\beq
\dot t = \frac{E}{f}\,,\qquad \dot \phi=\frac{L}{fr^2}\,.
\eeq
Plugging these relations into Eq. \eqref{null_met} one gets
\beq
{1\over f} \left(E^2 - \dot{r}^2 - {L^2\over r^2}\right) = 0 \,,
\eeq
namely
\beq
\label{eqdotr}
\dot{r}^2=E^2- {L^2\over r^2}=E^2 \left(1-\frac{b^2}{r^2} \right)\,,
\eeq
where $b=L/E$. Consequently, the orbit satisfies the equation
\beq 
d\phi = \pm {b dr \over f r^2 \sqrt{1- {b^2\over r^2}}}\,.
\eeq
Setting $u=M/r$ and introducing the adimensional quantity $\hat b={b}/{M}$ we find
\beq 
d\phi = \mp {\hat b  du \over \sqrt{(1-4u) (1- \hat b^2 u^2})}\,,
\eeq
which is an elliptic integral with three analytic roots: $u=1/4$ (cap/singularity) and 
$u=\pm 1/\hat b$ (turning points).
The critical case corresponds to $\hat b_{\rm crit}=4$ in which case
\beq
d\phi = \mp \frac{ 4 du}{(1-4u) \sqrt{1+4u} }\,.
\eeq
For $\hat b>4$ one has a scattering process and the scattering angle can be computed from the relation
\beq
\Delta\phi = \pi - \int_0^{1/\hat b} \frac{\hat b  du}{ \sqrt{(1-4u) (1- \hat b^2 u^2})}\,.
\eeq

In a scattering situation, it is often convenient to expand the integrand for large $\hat b$ and then integrate. This leads to 
\bea
\frac{\chi_{\rm null}+\pi}{2}&=&\frac{\pi}{2}+\frac{2}{\hat b}+\frac{3}{2 \hat b^2} \pi+\frac{40}{3\hat b^3}+\frac{105}{8\hat b^4}\pi\nonumber\\
&+& \frac{672}{5\hat b^5} +\frac{1155}{8\hat b^6}\pi+O\left(\frac{1}{\hat b^7} \right)\,.
\eea
A posteriori, this expression can be resummed as
\bea
&&\frac{\chi_{\rm null}+\pi}{2}=\frac{1}{(1-4 y)^{1/2}} \, K \left(\frac{-8 y }{ 1-4 y }\right)\nonumber\\ 
&+& 
 2 y\,  {}_3F_2  
[\{\frac34, 1, \frac54\}, \{\frac32, \frac32\}, 16 y^2]\,,
\eea
with $K$ a complete elliptic integral and
\beq
y= 
\frac{1}{\hat b}\,,
\eeq
which is of special interest for applications.

\begin{figure*}
\[
\begin{array}{cc}
\includegraphics[scale=0.75]{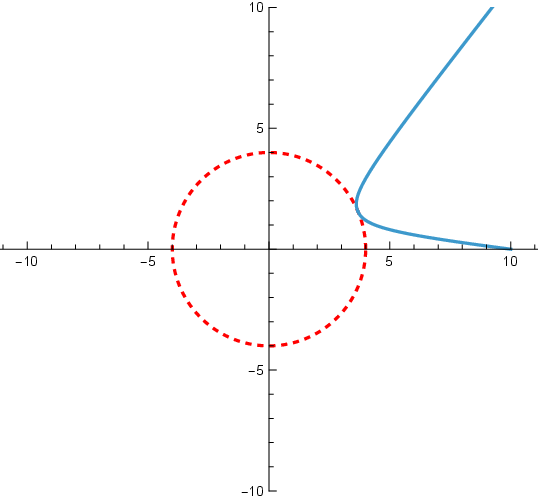}& \includegraphics[scale=0.75]{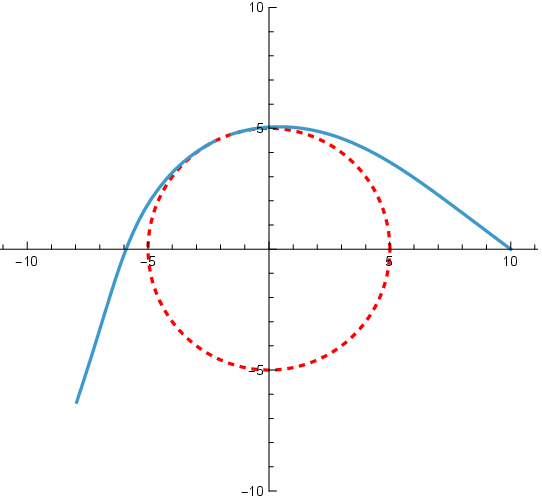}\\
(a) & (b)\\
\end{array}
\]
\caption{\label{fig:1} Null geodesics starting from $r(0)=10$, $\phi(0)=0$  for 
$\hat b=M$ (Panel (a)) and $\hat b=5M$ (Panel (b)).
In the first case ($r(0)>b$, $b<4M$) the orbit  approaches the naked singularity at $r=4M$ and it bounces back undergoing a scattering process. In the second case ($r(0)>b$, $b>4M$) the scattering process implies turning around the naked singularity. In both plots the mass parameter  has been chosen so that $M=1$. 
}
\end{figure*}

Noticeably, Eq. \eqref{eqdotr} can also be integrated exactly. To do this one introduces a new affine parameter $\sigma$ and set
\beq
\tau=\frac{b}{E} \sinh \sigma\,,
\eeq
so that
\beq
\frac{dr}{d\tau}=\frac{\frac{dr}{d\sigma}}{\frac{d\tau}{d\sigma}}=\frac{E }{b \cosh \sigma}\frac{dr}{ d\sigma}\,,
\eeq
and
\beq
\label{dr_dsigma}
\frac{E}{b \cosh \sigma} \frac{dr}{d\sigma} = E \sqrt{1-\frac{b^2}{r^2(\sigma)}}\,.
\eeq
The solution of Eq. \eqref{dr_dsigma} is then
\beq
r(\sigma)=b\cosh \sigma\,,\quad \frac{dr}{d\sigma}=b \sinh \sigma\,,
\eeq
that is
\beq
r(\tau)=b \sqrt{1+\left(\frac{E\tau}{b}\right)^2}\,.
\eeq 
Unfortunately, $\phi(\sigma)$ is not equally simple:
\beq
\frac{E}{b\cosh \sigma }\frac{d\phi}{d\sigma}=\frac{L}{b^2\cosh^2 \sigma}\frac{1}{\sqrt{1-\frac{4M}{b \cosh \sigma}}}\,,
\eeq
that is
\beq
\frac{d\phi}{d\sigma}=\frac{1}{\cosh  \sigma}\frac{1}{\sqrt{1-\frac{w}{\cosh \sigma}}}\,,
\eeq
where we have set
\beq
\label{w_def}
w=\frac{4M}{b}\,.
\eeq
However, the latter equation can be also solved exactly in terms of an incomplete  elliptic integral 
\begin{widetext}
\bea
\phi(\sigma)= \frac{2}{\sqrt{w+1}} 
F\left({\rm arcsin}\left(\frac{\sqrt{\cosh (\sigma )-w}}{\sqrt{w} \sqrt{\cosh (\sigma )-1}}\right)|\frac{2 w}{w+1}\right)+{\rm const}\,.
\eea

\end{widetext}
In the limit  $w\ll 1$, 
and choosing the integration constant such that $\phi(0)=0$, 
one finds
\bea
\phi(\sigma)&=&\left(2+\frac38 w^2 +\frac{105}{512} w^4\right)\left[{\rm arctan}(e^\sigma)-\frac{\pi}{4}\right] \nonumber\\
&+&\tanh \sigma 
\left[\frac12 w+\frac{3}{16\cosh(\sigma)}w^2\right.\nonumber\\
&+& \left(\frac{5}{ 48\cosh(\sigma)^2 }+\frac{5}{24}\right)w^3\nonumber\\
&+&\left. \left(\frac{105}{ 1024\cosh(\sigma)}+\frac{35}{ 512\cosh(\sigma)^3}\right)w^4\right]\nonumber\\
&+& O(w^5)\,.
\eea

Let us conclude this subsection with a comment.
Ref. \cite{Jantzen:1992rg} (see Eq. (12.8) there; see also Ref. \cite{Bini:2014rga} for applications) has discussed light propagation in a given spacetime as if its spatial geometry would act as an optical medium with a certain (observer-dependent) refraction index. For the one associated with the static observers (our fiducial observes) one has
\beq
n_{\rm (refr)}=\frac{1}{\sqrt{-g_{tt}}}=\frac{1}{f^{1/2}}=\left(1-\frac{4M}{r}\right)^{-1/4}\,,
\eeq
with $n_{\rm (refr)}>1$ (approaching $1$ only at spatial infinity and diverging at $r=4M$).
This reinforces the idea that the naked singularity should show repulsive behaviors once approached. In fact, light propagating towards the singularity meets an infinite refraction index (i.e., an optically infinitely dense medium) and is repelled.

\subsection{Timelike geodesics}

Also in the  case of timelike geodesics, because of the spherical symmetry one can study equatorial geodesics without loss of generality. As we will see later on, the bottom-up approach based on naive 4d description is misleading, since it neglects `warping' effects in the mass and Kaluza-Klein (KK) momentum. Notwithstanding this {\it caveat}, massive geodesics are described by the following equations
\bea
\frac{dt}{d\tau}&=& \frac{E}{f}\,, \qquad
\frac{d\phi}{d\tau}=  \frac{L}{r^2f}\,, \nonumber\\
\left(\frac{dr}{d\tau}\right)^2 &=&  E^2-\frac{L^2}{r^2} -f \mu^2\,,
\eea 
where $\mu$ denotes the 4d mass of the probe\footnote{As already mentioned, we are neglecting the `warping', that requires a full 5d description.} and $\tau$ its proper time.
Note that the region for allowed motion is 
\beq
 E^2-\frac{L^2}{r^2}\ge f \mu^2\,.
\eeq
Circular timelike critical geodesics exist at any radius $r_0>4M$ and are characterized by
\bea
E^2=\frac{(1-3u_0)^2}{\sqrt{1-4u_0}}\,,\quad \frac{L^2}{M^2}=\frac{1}{u_0 \sqrt{1-4u_0}}\,,
\eea
where $u_0=M/r_0$, as follows by solving the equations ${d^2 r}/{d\tau^2}=0={dr}/{d\tau}$.

The scattering angle for a massive probe reads instead
\beq
\frac{\chi+\pi}{2}=\int_\infty^{r_1}\frac{J}{rf \sqrt{r^2E^2-J^2-r^2f \mu^2}}
\eeq
Following \cite{Damour:1988mr,Ivanov:2025ozg,Bini:2025ltr,Bini:2025bll}, the `partie finie' of the integral is captured by expanding the external turning point $r_1$ in the upper end of integration at first order in Post-Minkowskian (PM) expansion
\beq
r^{(0)}_1=\frac{J}{\sqrt{E^2-\mu^2}}
\eeq
If we change radial variable to $u={r^{(0)}_1}/{r}$ and introduce the notation 
\bea
\label{notation_J}
J&=& jM m_4 =b \sqrt{E^2-\mu^2}=b \mu \sqrt{\gamma^2-1}=b \mu p_\infty\,,\nonumber\\
y&=&\frac{p_\infty}{j}=\frac{1}{\hat b}\,,
\eea
the above integral becomes
\beq
\frac{\chi+\pi}{2}=\int_{0}^1\frac{du}{\sqrt{1-4u y}\sqrt{1-u^2}\sqrt{1-\frac{\sqrt{1-4 u y}-1}{p_\infty^2(1-u^2)}}}\,,
\eeq
which can be expanded as follows
\bea
\frac{\chi+\pi}{2}&=&\sum_{k=0}^\infty\int_0^1\frac{du}{\sqrt{1-u^2}\sqrt{1-4uy}}\times \nonumber\\
&& \binom{-\frac{1}{2}}{k}\left(\frac{1-\sqrt{1-4uy}}{p_\infty^2(1-u^2)}\right)^k\,,
\eea
taking into account
\bea
&&\frac{(1{-}\sqrt{1{-}4uy})^k}{\sqrt{1{-}4uy}}\equiv (2uy)^k{}_2F_1\left(\frac{1{+}k}{2},\frac{2{+}k}{2},1{+}k,4u y\right)\nonumber\\
&&=\sum_{n=0}^\infty\frac{2^{k{+}2 n{-}2} \left(\frac{k}{2}{+}1\right)_{n{-}1}
   \left(\frac{k{+}1}{2}\right)_{n{-}1}}{(1)_{n{-}1} (k{+}1)_{n{-}1}}(u y)^{n{+}k{-}1}\,.
\eea
After integration and resummation on the index $n$, we obtain
\begin{widetext}
    \bea
\frac{\chi+\pi}{2}&=&\frac{\pi^{3/2}}{2}\sum_{k=0}^\infty\frac{y^k}{p_\infty^{2k}}\Big[ 
   {}_4\tilde{F}_3\left(\frac{k+1}{4},\frac{k+2}{4},\frac{k+3}{4},\frac{k+4
   }{4};\frac{1}{2},1-\frac{k}{2},\frac{k+2}{2};16 y^2\right)\nonumber\\
   &+&\frac14 (k+1) (k+2)
   y \,
   _4\tilde{F}_3\left(\frac{k+3}{4},\frac{k+4}{4},\frac{k+5}{4},\frac{k+6
   }{4};\frac{3}{2},\frac{3-k}{2},\frac{k+3}{2};16 y^2\right)\Big]\,,
    \eea
\end{widetext}
in terms of regularized hypergeometric functions, denoted by a tilde.
As before, the scattering angle can be computed in a large $\hat b$ expansion and turns out to be given by
\bea
\frac{\chi+\pi}{2} &=&  p_\infty^0 f_0(y)+ \frac{1}{p_\infty^2}f_{-2}(y)\nonumber\\
&+&\frac{1}{p_\infty^4}f_{-4}(y)+ \frac{1}{p_\infty^6}f_{-6}(y)+...\,,
\eea
where
\beq
f_0(y)=\frac{\chi_{\rm null}+\pi}{2}\,,
\eeq
and, for example
\bea
f_{-2}(y)&=&\frac{3}{2} \pi  y^2 \, _2F_1\left(\frac{5}{4},\frac{7}{4};2;16 y^2\right)\nonumber\\
&+& y \, _3F_2\left(\frac{3}{4},1,\frac{5}{4};\frac{1}{2},\frac{3}{2};16 y^2\right)\,.
\eea
For completeness, the scattering angle in full expanded form is instead given by
\begin{widetext}
\begin{eqnarray}
\label{chi_4d}
 \chi  &=&  
y^8\frac{3465 \pi   \left(65 p_{\infty }^8 + 260 p_{\infty }^6 + 364 p_{\infty }^4 + 208 p_{\infty }^2 + 40\right)}{64 p_{\infty }^8} \nonumber\\
&+& y^7 \frac{2 \left(54912 p_{\infty }^{14} + 192192 p_{\infty }^{12} + 221760 p_{\infty }^{10} + 92400 p_{\infty }^8 + 8400 p_{\infty }^6 - 504 p_{\infty }^4 + 56 p_{\infty }^2 - 5\right)}{35 p_{\infty }^{14}} \nonumber\\
&+&y^6 \frac{105 \pi   \left(11 p_{\infty }^6 + 33 p_{\infty }^4 + 30 p_{\infty }^2 + 8\right)}{4 p_{\infty }^6} \nonumber\\
&+& y^5 \left(
\frac{672}{p_{\infty }^2} + \frac{448}{p_{\infty }^4} + \frac{56}{p_{\infty }^6}
- \frac{4}{p_{\infty }^8} + \frac{2}{5 p_{\infty }^{10}} + \frac{1344}{5}
\right) \nonumber\\
&+&  y^4 \frac{15 \pi \left(7 p_{\infty }^4 + 14 p_{\infty }^2 + 6\right)}{4 p_{\infty }^4} \nonumber\\
&+& y^3 \left(
\frac{40}{p_{\infty }^2} + \frac{8}{p_{\infty }^4} - \frac{2}{3 p_{\infty }^6} + \frac{80}{3}
\right) \nonumber\\
&+& 3 \pi y^2 \left(\frac{1}{p_{\infty }^2} + 1\right) \nonumber\\
&+& y \left(\frac{2}{p_{\infty }^2} + 4\right)\,.
\end{eqnarray}

\end{widetext}

\subsection{Newmann-Penrose (NP) formalism}

The neutral $W$-soliton spacetime in 4d is of type D, and the two repeated principal null directions (PNDs) are
\bea
K^\pm_{\rm PND} =e_{\hat t} \pm e_{\hat r}\,,
\eea
where
\bea
e_{\hat t}&=& \frac{1}{\sqrt{f}}\partial_t \,,\qquad
e_{\hat r}=\sqrt{f} \partial_r \nonumber\\
e_{\hat \theta} &=& \frac{1}{r\sqrt{f}} \partial_\theta \,,\qquad
e_{\hat \phi}= \frac{1}{r\sin \theta \sqrt{f}} \partial_\phi\,,
\eea
With respect to a natural NP frame 
\bea
\label{NP_tetrad}
l&=&\frac{\sqrt{f}}{\sqrt{2}}(e_{\hat t}+e_{\hat r})\,,\qquad
n=\frac{1}{\sqrt{f}\sqrt{2}}(e_{\hat t}+e_{\hat r})\nonumber\\
m&=& \frac{1}{\sqrt{2}}(e_{\hat \theta}+i e_{\hat \phi})\,,
\eea
the nonvanishing spin coefficients are given by
\bea
\mu &=&\frac{1}{\sqrt{2}} \frac{(3M-r)}{r^2f^2}=\frac{1}{f}\rho\,,\nonumber\\
\alpha &=& -\frac{1}{2\sqrt{2}} \frac{\cot \theta}{r\sqrt{f}}=-\beta\,,\nonumber\\
\epsilon &=& \frac{1}{\sqrt{2}}  \frac{M}{ f r^2}\,,
\eea
and
the only non-vanishing Weyl scalar is
\beq
\label{psi_2}
\Psi_2=-\frac{ M }{f r^3}
\eeq
implying that the spacetime is of Petrov type D, as already stated above. 

Yet, contrary to asymptotically flat BHs in 4d or Myers-Perry BHs,  in the present case it is not possible to achieve a Kerr-Schild form for the metric since the PNDs of the metric differ from the PNDs of the energy momentum of the scalar field source of the spacetime. Actually,  the tetrad \eqref{NP_tetrad} allows for a single non-vanishing Weyl tensor component \eqref{psi_2} but the 
energy momentum tensor of the complex scalar field is represented by three non-vanishing Ricci coefficients  
\bea
[\Phi_{00},\Phi_{11},\Phi_{22}]=\frac{3M^2}{2r^4 f^2}\left[1,-\frac{1}{2f},\frac{1}{f^2}  \right]\,,
\eea
meaning misalignment of the PNDs. To make a long story short, 
the tetrad  \lq\lq diagonalizing" the Weyl tensor does not reduce to \lq\lq diagonal form" the energy momentum tensor of the source.
Let us point out that this is a crucial difference with the black hole case. For example, for the Reissner-Nordstr\"om spacetime (in standard coordinates) a natural tetrad like the one in Eq. \eqref{NP_tetrad} implies only two nonvanishing components
\beq
\Psi_2^{\rm RN}=\frac{Q^2-Mr}{r^4}\,,\qquad \Phi_{11}=\frac{Q^2}{2r^4}\,,
\eeq
i.e., the above mentioned \lq\lq alignment."

We now pass to examine perturbations, starting from the simplest case of source-free scalar perturbations. 

\subsection{Non minimal coupling of the scalar field}

The source-free massive scalar field equation, in the presence of non-minimal coupling to the scalar curvature,  is represented by the equation 
\beq
\label{waveeq}
\left(\Box-\mu^2-\xi R\right)\phi=0\,.
\eeq
Two values of $\xi$ are of particular interest. The minimally coupled case $\xi=0$ and the conformally coupled case ($D$ is the spacetime dimension)
\be
\xi=\frac{D-2}{4(D-1)}\underset{D\to 4}{\longrightarrow}\frac{1}{6}\,.
\ee
The wave equation \eqref{waveeq} can be reduced to a single ODE by expanding in spherical harmonics and Fourier transforming the time
\be
\phi(t,r,\theta,\phi)=\sum_{lm} Y_{\ell m}(\theta, \phi)\int \frac{d\omega}{2\pi}e^{-i\omega t }R_{\ell \omega}(r)\,,
\ee
so that  
\bea\label{eqradial}
&&r(r-4M)R_{\ell \omega}''(r)+2(r-2M)R_{\ell \omega}'(r)\nonumber\\
&+&\Big[r^2\omega^2-\ell (\ell+1)-\frac{6M^2\xi}{r(r-4M)}\nonumber\\
&-&r^2\mu^2\sqrt{1-\frac{4M}{r}}\Big]R_{\ell \omega}(r)=0\,.
\eea
is actually $m$ independent, thanks to spherical symmetry.

The latter equation \eqref{eqradial} can be cast in its normal  form via the following rescaling (denoting $R_{\ell\omega}$ by $R$ for brevity)
\be
R(r)=\frac{\psi(r)}{rf}\,,
\ee
so that it becomes
\beq
\label{psi_eq}
\psi''(r)+Q_{\rm W}(r)\psi(r)=0\,,
\eeq
where
\bea
\label{Q_W_def}
Q_{\rm W}(r)&=&\frac{r(r-4M)(r^2\omega^2-\ell (\ell +1))+2M^2(2-3\xi)}{r^2(r-4M)^2}\nonumber\\
&-&\frac{\mu^2}{\sqrt{1-\frac{4M}{r}}}\,.
\eea
The presence of a mass term $\mu$ in Eq. \eqref{Q_W_def}  yields a non-Fuchsian term in the wave equation. For this reason, hereafter we will focus only on the case $\mu=0$ (still keeping $\xi \not=0$). This situation is described by a Confluent Heun Equation (CHE) or, equivalently, it can be mapped to the qSW  curve with $N_f=3$ flavors (in the Hanany-Witten setup with 1 flavor brane on the Left and 2 on the right) \cite{Aminov:2020yma,Aminov:2023jve,Bautista:2023sdf,Bonelli:2022ten,Bonelli:2021uvf,Bianchi:2021xpr,Bianchi:2021mft,Consoli:2022eey,Fucito:2023afe,Cipriani:2024ygw,Bianchi:2023rlt} so that 
\bea\label{Q12}
Q_{1,2}(y)&=&{-}\frac{q^2}{4}{+}\frac{1{-}(m_1{-}m_2)^2}{4y^2}{+}\frac{1{-}(m_1{+}m_2)^2}{4(1{+}y)^2}\nonumber\\
&{-}&\frac{m_3 q}{y}{+}\frac{2(m_1^2{+}m_2^2){-}1{-}4u{+}2q(1{-}m_1{-}m_2)}{4y(1{+}y)}\,,\nonumber\\
\eea
where
\be
y=\frac{r-4M}{4M}\,.
\ee
The dictionary for the `gauge-gravity' correspondence $Q_{\rm W} \leftrightarrow Q_{1,2}$  is the following
\bea
\label{dictionary}
q&=&8iM\omega,\quad m_1=0,\quad m_2=-\sqrt{\frac{3\xi}{2}},\quad m_3=2iM \omega\,,\nonumber\\
u&=&\left(\ell+\frac{1}{2}\right)^2+2iM\omega(2+\sqrt{6\xi})\,.
\eea
In the minimally coupled case  ($\xi=0$) it simplifies to
\bea
q&=& 8iM\omega,\qquad m_1=m_2=0,\quad m_3=2iM \omega\nonumber\\
u&=&\left(\ell+\frac{1}{2}\right)^2+4iM\omega\,.
\eea
This means that in the decoupling limit $q=0$, the solution of the differential equation in normal form with effective potential \eqref{Q12} is
\bea
\psi_0(r)&=&\sqrt{y(1+y)}\Big[c_1 P_{\nu}\left(1+2y\right)\nonumber\\
&+&c_2 Q_{\nu}\left(1+2y\right)\Big]
\eea
where $c_{1,2}$ are integration constants, $\nu=\sqrt{u}-\frac{1}{2}$ and $P_\nu=P_\nu^{\mu=0}$ and $Q_\nu=Q_\nu^{\mu=0}$ are the Legendre (associated) functions describing respectively the regular and the irregular behaviour at $y=0$ or, equivalently, at $r=4M$.
We will consider now  JWKB-type of solutions, Leaver  \lq\lq continuous fraction" solutions, and  numerical integrations to compute the corresponding scalar QNMs.

\subsubsection{JWKB-type of solutions}

The (adapted) Born-Sommerfeld quantization condition reads
\beq
\label{quant_cond}
\int_{r_-}^{r_+}\sqrt{Q_{\rm W}(r)}dr=\pi\left(n+\frac{1}{2}-\frac{m_2}{2}\right)\,,
\eeq
where $n$ is a non-negative integer called the overtone number and $r_{\pm}$ are the classical turning points. 
The presence of the shift due to $m_2$ in Eq. \eqref{quant_cond} is justified below (by using the qSW curve for convenience).

Since $m_1=0$ from the dictionary \eqref{dictionary}, in the decoupling limit ($q=0$) the qSW curve \eqref{Q12} reduces to
\bea\label{Q12bis}
Q_{1,2}^{q=0}(y)&=&\frac{1-m_2^2}{4y^2}+\frac{1-m_2^2}{4(1+y)^2}+\frac{2m_2^2-1-4u}{4y(1+y)}\,.\qquad 
\eea
The conditions for the critical regime read
\beq
Q_{1,2}^{q=0}(y)=0=\frac{d}{dy} Q_{1,2}^{q=0}(y)\,,
\eeq
and are solved by
\beq
\label{y_sol}
y=-\frac12\,, \qquad u=m_2^2 -\frac34\,.
\eeq
To recover the JWKB region of validity one has to assume $u\propto \ell^2 \gg 1$ and hence $m_2\propto \ell\gg 1$ from Eq. \eqref{y_sol}.
This implies 
\beq
\frac{Q_{1,2}^{q=0}(y)}{\sqrt{2 \frac{d^2}{dy^2}Q_{1,2}^{q=0}(y)}}\Bigg|_{y=-\frac12}=\frac{i}{2}\sqrt{u}- \frac{i}{2}m_2\,,
\eeq
which leads to the condition \eqref{quant_cond}.

In the semiclassical limit, the turning points coalesce, and the integrand can be approximated by
\bea
\int_{r_-}^{r_+}\sqrt{Q_{\rm W}(r)}dr&\sim&\int_{r_-}^{r_+}\sqrt{Q_{\rm W}(r_c)+\frac{Q_{\rm W}''(r_c)}{2}(r-r_c)^2}dr\nonumber\\
&\sim&\frac{i\pi Q_{\rm W}(r_c)}{\sqrt{2Q_{\rm W}''(r_c)}}\,,
\eea
where $r_c\in [r_-,r_+]$ is the critical point of the effective potential, i.e. $Q_{\rm W}(r_c,\omega_c)=Q_{\rm W}'(r_c,\omega_c)=0$. 
Finally, one finds
\beq
\label{eq268}
\frac{Q_{\rm W}(r_c,\omega)}{\sqrt{2Q_W''(r_c,\omega)}}=-i\left(n+\frac{1}{2}\right)+\frac{im_2}{2}\,.
\eeq 
The previous quantization condition \eqref{quant_cond} can be solved perturbatively plugging $\omega_{\rm JWKB}=\omega_R+i \omega_I$ in Eq. \eqref{eq268}, and expanding for small $\omega_I$. In the eikonal limit ($\ell \gg 1$),   the following relations hold
\bea
\frac{r_c}{M}&{=}&4+\frac{\sqrt{2{-}3\xi}}{\ell}{+}O \left(\frac{1}{\ell^2}\right)\,, \nonumber\\
M \omega_R&=&M \omega_c=\frac{\ell}{4}{+}\frac{1{-}\sqrt{2{-}3\xi}}{8}{+}O \left(\frac{1}{\ell}\right)\,, \nonumber\\
M \omega_I&=&-\frac{4 n+\sqrt{6\xi
   }+2}{8 \sqrt{2} }\Big[1 + \frac{1}{\ell}\nonumber\\
& \times &\left(\frac{3}{16}\frac{\xi+2}{\sqrt{2-3\xi}}-1 \right)\Bigg] 
+O\left(\frac{1}{\ell^2}\right)\,. 
\eea
These results (provided by the JWKB approach) will be used below in Table \ref{TAB1} and in Fig. \ref{fig1}.

\subsubsection{Leaver continuous fraction}

The dominant behavior at $r=4M$ is described by the ansatz
\be
R(r)\underset{r\to4M}{\sim}(r-4M)^{\delta_\pm}\,,
\ee
with the two Frobenius exponents given by
\be
\delta_\pm=\pm\frac{1}{2}\sqrt{\frac{3\xi}{2}}\,.
\ee
It is worth noticing that in the minimally coupled case ($\xi=0$) the wave ends smoothly in $r=4M$, while for generic $\xi$ the exponent with the plus (minus) sign describes the (ir)regular behavior at the singularity.
At infinity the singularity is irregular, and the leading behavior is
\be
R(r)\underset{r\to\infty}{\sim} e^{\pm i \omega r}
\ee
describing the ingoing (\lq\lq-") and outgoing (\lq\lq +") waves at infinity.

QNMs are defined by imposing the conditions of having outgoing waves at infinity and regularity at the `cap' (the naked singularity from a strictly 4d perspective)\footnote{For BHs, one has to impose ingoing boundary conditions at the horizon.}. In particular, let's consider the following ansatz for \eqref{eqradial}
\be
\label{sol_leav}
R(r)=e^{i\omega r}(r-4M)^{\frac{1}{2}\sqrt{\frac{3\xi}{2}}}r^\sigma \sum_{n=0}^\infty c_n\left(\frac{r-4M}{r}\right)^n\,,
\ee
where   the exponent $\sigma$ can be chosen freely, since no physical boundary condition should be imposed at $r=0$. One possibility is to choose $\sigma$  in order to minimize the number of terms in the recursion relation which follows when inserting the solution of the form \eqref{sol_leav} in Eq. \eqref{eqradial}. 
The terms in the recursion become three if  
\beq
\label{sigma_def}
\sigma= -1-\frac{\sqrt{6\xi}}{4}+2iM\omega \,.
\eeq
With this choice we find the following structure
\be
\sum_{n=0}^\infty c_n\Big[\frac{A_n}{z}+B_n+C_n z\Big]z^n=0\,.
\ee
We can redefine the index of the sum in order to obtain
\be
\sum_{n=0}^\infty \Big[c_{n+1}A_{n+1}+c_nB_n+c_{n-1}C_{n-1} \Big]z^n=0\,,
\ee
setting $C_{-1}=0$. Finally, the recursion relation to solve reduces to the following three-terms relation
\begin{widetext}
\bea
\alpha_n&=&A_{n+1}=\frac{1}{2}(n+1)(2+2n+\sqrt{6\xi})\,,\nonumber\\
\beta_n&=&B_n=-\ell(\ell +1)-1-2n^2-n \left(-12 i M \omega +\sqrt{6\xi}
    +2\right)+\sqrt{\frac{3}{2}}
   \sqrt{\xi } (-1+6 i M \omega
   )+2 M \omega  (8 M \omega +3
   i)\,,\nonumber\\
\gamma_n&=&C_{n-1}=\frac{1}{2}(n-2iM\omega)(2n+\sqrt{6\xi}-4iM\omega)\,.
\eea
\end{widetext}
This recursion relation is solved in terms of the continuous fractions 
\be
\label{fractionLeaver}
\beta_n =\frac{\alpha_n\gamma_n}{\beta_{n-1}-\frac{\alpha_{n-2}\gamma_{n-1}}{\beta_{n-2}-\dots}}+\frac{\alpha_n\gamma_{n+1}}{\beta_{n+1}-\frac{\alpha_{n+1}\gamma_{n+2}}{\beta_{n+2}-\dots}}\,.
\ee
We choose the continuous fraction with $n=0$, and solve it numerically in $\omega$, providing an estimate for QNM frequencies.

\subsubsection{Numerical integration}

Let us briefly describe the numerical procedure used to compute QNMs frequencies. The starting point consists in finding the leading and a sufficient number of subleading behaviours at infinity and at the singularity. 
At infinity, the radial wave function behaves as 
\beq\label{Rinf}
R_\infty(r)\underset{r\to \infty}{\sim} e^{i\omega r}r^{-1+2iM\omega}\sum_{n=0}^{N_\infty}k_n r^{-n}\,,
\eeq
where
\bea
M \frac{k_1}{k_0}&=& 1 -6i \hat \omega +\frac{iL}{2\hat \omega}\,,\nonumber\\
M^2 \frac{k_2}{k_0}&=& \frac{5}{2} +\frac{i}{2\hat \omega}-16 i \hat \omega -18 \hat \omega^2\nonumber\\ 
&+& L \left(3+\frac{i}{\hat \omega}+\frac{1}{4\hat \omega^2} \right)
-\frac{L^2}{8\hat \omega^2}\,,
 \eea
etc., where we used the notation $L=\ell(\ell+1)$, $\hat \omega=M\omega$.

The correct behaviour at the naked singularity requires regularity of the wave
\bea\label{Rhot}
R_{\rm NS}(r)&\underset{r\to4M}{\sim}& (r-4M)^{\frac{1}{2}\sqrt{\frac{3\xi}{2}}}\sum_{n=0}^{N_{\rm NS}}\tilde{k}_n (r-4M)^n\nonumber\\
&\underset{\xi\to \frac16}{\to}& \quad (r-4M)^ \frac{1}{4}\sum_{n=0}^{N_{\rm NS}}\tilde{k}_n (r-4M)^n\,,
\eea
and the first coefficients are listed below (for $\xi=\frac16$)
\bea
M \frac{\tilde k_1}{\tilde k_0}&=& -\frac{1}{16}-\frac{8\hat \omega^2}{3}+\frac{L}{6}\,,\nonumber\\
M^2 \frac{\tilde k_2}{\tilde k_0}&=&\frac{5}{512}+\frac{\hat \omega^2}{30}+\frac{32\hat \omega^4}{15}\nonumber\\
&-& L \left(\frac{13}{480}+\frac{4\hat \omega^2}{15} \right)
+\frac{L^2}{120}\,.
\eea

\begin{figure*}
    \centering
    \includegraphics[width=0.6\linewidth]{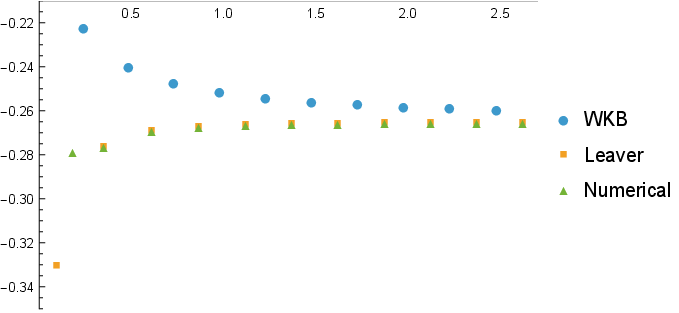}
    \caption{Lowest overtone QNMs for the conformally coupled case $\xi=1/6$. $M$ is set to be 1.}
    \label{fig1}
\end{figure*}

The numerical integration is performed starting from the naked singularity $r=4M+\epsilon$ ($\epsilon=10^{-3}$) and proceeding up to an extraction point $r_*$ (we choose $r_*=8$), and from this middle-point to infinity (set at $r=25$ in some units). The boundary conditions of the two numerical integrations are respectively \eqref{Rhot} and \eqref{Rinf} and their derivatives. We are considering here $N_{\rm NS}=N_\infty=25$ subleading terms at the siungularity and at infinity. The numerical Wronskian can be then constructed,  and its zeros are the QNMs frequencies. Comparison among JWKB, Leaver continuous fraction and numerical method is shown in Tab. \ref{TAB1xi0}, \ref{TAB1} and Fig. \ref{fig1}.

    \begin{table*}
\begin{tabular}{|c|c|c|c|}
\hline
$\ell$ & JWKB                  & Leaver               & Numerical Integration                   \\ \hline
$0$ & $-$ & $0.136693 - 0.205453 {\rm i}$ & $0.102588 - 0.21277 {\rm i}$ \\ \hline
$1$ & $0.22481 - 0.14396 {\rm i}$ & $0.375423 - 0.181936{\rm i} $ & $0.37543 - 0.181932{\rm i} $ \\ \hline
$2$ & $0.465385 - 0.157496 {\rm i}$ & $0.62508 - 0.178715 {\rm i}$ & $0.62508 - 0.178715{\rm i} $ \\ \hline
$3$ & $0.710836 - 0.163115{\rm i} $ & $0.875028 - 0.177778{\rm i} $ & $0.875028 - 0.177778{\rm i} $ \\ \hline
$4$ & $0.95818 - 0.166197{\rm i} $ & $1.12501 - 0.177385{\rm i} $ & $1.12501 - 0.177385{\rm i} $ \\ \hline
$5$ & $1.20644 - 0.168145{\rm i} $ & $1.37501 - 0.177185{\rm i} $ & $1.375 - 0.177176{\rm i} $ \\ \hline
$6$ & $1.45522 - 0.169487{\rm i} $ & $1.625 - 0.17707{\rm i} $ & $1.6248 - 0.177056{\rm i} $ \\ \hline
$7$ & $1.70431 - 0.170468{\rm i} $ & $1.875 - 0.176997{\rm i} $ & $1.875 - 0.176997{\rm i} $ \\ \hline
$8$ & $1.95362 - 0.171216 {\rm i}$ & $2.125 - 0.176948{\rm i} $ & $2.125 - 0.176948{\rm i} $ \\ \hline
$9$ & $2.20306 - 0.171805{\rm i} $ & $2.375 - 0.176914{\rm i} $ & $2.375 - 0.176914{\rm i} $ \\ \hline
$10$ & $2.45261 - 0.172282 {\rm i}$ & $2.625 - 0.176889 {\rm i}$ & $2.625 - 0.176889 {\rm i}$ \\ \hline
\end{tabular}
\caption{\label{TAB1xi0}Lowest overtone QNMs for the minimally coupled case $\xi=0$ and in units of $M=1$.  The JWKB values here follow from direct (numerical) use of Eq. \eqref{eq268}. They differ from the corresponding column in Table I of  Ref. \cite{Dima:2025tjz} which was using instead the \lq\lq eikonal limit" formula
$M\omega=\frac{1}{4}(\ell+\frac12)-\frac{i}{2\sqrt{2}}(n+\frac12)$.}
\end{table*}
\begin{table*}
\begin{tabular}{|c|c|c|c|}
\hline
$\ell$ & JWKB                  & Leaver               & Numerical Integration                   \\ \hline
$0$ & $-$ & $0.0929455 - 0.330645 {\rm i}$ & $0.178016 - 0.27855 {\rm i}$ \\ \hline
$1$ & $0.237356 - 0.222225 {\rm i}$ & $0.34818 - 0.276139 {\rm i}$ & $0.347894 - 0.276195 {\rm i}$ \\ \hline
$2$ & $0.481882 - 0.240086 {\rm i}$ & $0.608171 - 0.269093 {\rm i}$ & $0.608171 - 0.269093 {\rm i}$ \\ \hline
$3$ & $0.72924 - 0.247419 {\rm i}$ & $0.86285 - 0.267162 {\rm i}$ & $0.86285 - 0.267162 {\rm i}$ \\ \hline
$4$ & $0.977698 - 0.251428 {\rm i}$ & $1.11551 - 0.266371 {\rm i}$ & $1.11551 - 0.266371 {\rm i}$ \\ \hline
$5$ & $1.22669 - 0.253959 {\rm i}$ & $1.36722 - 0.265971 {\rm i}$ & $1.36722 - 0.265972 {\rm i}$ \\ \hline
$6$ & $1.47598 - 0.255701 {\rm i}$ & $1.61841 - 0.265742 {\rm i}$ & $1.61841 - 0.265742 {\rm i}$ \\ \hline
$7$ & $1.72545 - 0.256975 {\rm i}$ & $1.86928 - 0.265598 {\rm i}$ & $1.86928 - 0.265598 {\rm i}$ \\ \hline
$8$ & $1.97505 - 0.257946 {\rm i}$ & $2.11995 - 0.265502 {\rm i}$ & $2.11995 - 0.265502 {\rm i}$ \\ \hline
$9$ & $2.22472 - 0.258711 {\rm i}$ & $2.37048 - 0.265435 {\rm i}$ & $2.37048 - 0.265435 {\rm i}$ \\ \hline
$10$ & $2.47446 - 0.25933 {\rm i}$ & $2.62091 - 0.265386 {\rm i}$ & $2.62091 - 0.265386 {\rm i}$ \\ \hline
\end{tabular}
\caption{\label{TAB1}Lowest overtone QNMs for the conformally coupled case $\xi=1/6$   and in units of $M=1$.} 
\end{table*}

\section{5d neutral W soliton}

Let us pass now to analyze the genuine 5d neutral $W$-soliton solution \cite{Chakraborty:2025ger, Dima:2025tjz}.
The metric reads
\bea
ds^2&=&{-}f_s(r) dt^2{+}\frac{f_s(r) dr^2}{2f_s(r){-}1}{+}f_s(r)r^2d\Omega_2^2\nonumber\\
&+&\frac{(2f_s(r){-}1)}{f_s(r)^2}dy^2\,,
\eea
where 
\beq
f_s(r)=1-\frac{2M}{r}\, \quad {\rm so \, that} \quad 2f_s(r){-}1=1-\frac{4M}{r}
\eeq
and the gauge field is given\footnote{At variant with respect to to \cite{Chakraborty:2025ger, Dima:2025tjz},  we use $y$ in place of $\psi$  to denote the coordinate of the extra `warped' circle.}
\beq
A=\frac{2M}{r-2M}dy \,.
\eeq
The components $A_y$ and $g_{yy}$ are related to the complex scalar field in the 4d solution after reduction along $y$. 

The solution is completely smooth and horizonless and thus `solitonic'. The geometry is capped at $r\ge 4M$. This can be seen changing radial coordinate to $\rho = \sqrt{r-4M} R_y/ 2\sqrt{2}M^{3/2}$  around $r\approx 4M$. Absence of conical singularity requires a $R_y=2\sqrt{2}M$, preventing scale separation between the soliton mass and the KK radius. Yet, accepting `mild' ${\bf Z}_k$ orbifold singularities, one finds $R_y=2\sqrt{2}M/k$\footnote{In the charged case, the situation is slightly more involved.} which che be made parametrically small. \cite{Chakraborty:2025ger, Dima:2025tjz}. 

Let us study the geodesics by using the Hamiltonian form.
The mass-shell condition is
\beq
\mathcal{H}=-\mu_5^2,\quad \mathcal{H}=g^{\mu\nu}P_\mu P_\nu\,,
\eeq
where $\mu_5$ denotes the 5d mass of the probe and the canonical conjugate momenta are defined as
\beq
P_\mu=\frac{\partial \mathcal{L}}{\partial \dot x^\mu}\,.
\eeq
Here a dot means differentiation with respect to an affine parameter $\lambda$. 
The conserved momenta are then given by
\bea
\label{momenta}
P_t&=& -E=-f_s(r) \dot{t}, \quad P_\phi\equiv J=f_s(r)r^2\sin(\theta^2)\dot{\phi}\,,\nonumber\\
P_y&=&\frac{2f_s(r)-1}{f_s(r)^2}\dot{y}\,.
\eea
Passing to the proper time $\tau=\mu_5 \lambda$,   Eqs. \eqref{momenta} have the same form with $E\to \hat E=E/\mu_5$, $P_\phi\equiv J\to \hat P_\phi=P_\phi/\mu_5$ and $\dot x^\alpha$ becoming $\frac{dx^\alpha}{d\tau}$.
It is convenient to introduce the dimensionless angular momentum variable $j=J/(M\mu_5)$.

Hereafter we will use the proper time as a parameter, set $P_y=0$ for simplicity, and, without loss of generality, study  equatorial motion
\be
P_\theta=0,\quad \theta=\frac{\pi}{2}\,.
\ee
The mass shell condition implies
\bea\label{pR5d}
\hat P_r^2&=&\frac{Q_R}{\mu_5^2}=\frac{(\hat E^2-1)r^2+2M  r- M^2 j^2 }{r(r-4M)}\nonumber\\
&=&\frac{(\hat E^2-1)(r-r_1)(r-r_2)}{r(r-4M)}\,,
\eea
where 
\beq
\label{r12sol}
r_{1,2}=M \frac{-1 \pm  \sqrt{j^2(\hat E^2-1)+1}}{ (\hat E^2-1)}
\eeq
are such that $r_1>4M>0>r_2$.
 
Eq. \eqref{pR5d} can be rewritten as
\beq
\dot{r}^2=\frac{(r-4M)[(E^2-\mu_5^2)(r^2-b^2)+2M\mu_5^2 r]}{r(r-2M)^2}\,,
\eeq
where we have introduced the impact parameter
\be
b=\frac{J}{\sqrt{E^2-\mu_5^2}}=\frac{M j}{\sqrt{\hat E^2-1}}\,.
\ee
The critical and circular geodesics are defined by the condition
\beq
\dot{r}=0=\ddot{r}\,,
\eeq
which is solved by
\beq
\label{hatE_and_j}
\hat E=\frac{E}{\mu_5}=\sqrt{1-\frac{M}{r}}\,,\qquad  j=\frac{J}{M \mu_5}=\pm \sqrt{\frac{r}{M}}\,.
\eeq

Notice that at $r=4M$ one finds $\dot r=0=\ddot r$ also for the values
\beq
r_c=4M,\quad j_c=2 \sqrt{4\hat{E}^2-2}\,,
\eeq
which coincide with solutions \eqref{hatE_and_j} in the limit  $r\to 4M$ and for $\hat E=\sqrt{3}/{2}$.

Timelike circular orbits in the $(t,\phi)$ pseudo 2-plane correspond to the four velocity 
\beq
U_{\rm circ}=\Gamma (\partial_t +\Omega \partial_\phi)\,,
\eeq
with
\beq
\Gamma=\frac{\sqrt{1-\frac{M}{r}}}{1-\frac{2M}{r}}\,,\qquad \Omega=\sqrt{\frac{M}{r^3-Mr^2}}\,,
\eeq
showing that Kepler law $\Omega_{\rm K}=\sqrt{\frac{M}{r^3}}$ is valid only asymptotically.

\subsection{Scattering angle}
Starting from
\beq
\dot{\phi}=\frac{J}{r(r-2M)},\quad \dot{r}=\frac{P_r(r-4M)}{r-2M}\,,
\eeq
the scattering angle is defined as
\be
\chi=2\phi(\infty,r_1)-\pi,\quad \phi(\infty,r_1)=\int_{\infty}^{r_1}\frac{\dot{\phi}}{\dot{r}}dr\,.
\ee
Introducing the dimensionless variable $u=r_1/r$ we have
\be\label{elliptic}
\phi(\infty,r_1)=\int_0^{1}\frac{b du}{\sqrt{(1-u)(r_1-4M u)(r_1-r_2 u)}}\,.
\ee
The previous integral can be expanded in Post-Minkowskian sense (i.e., large $\hat b$), so that the final result reads
\begin{widetext}
\bea
\chi&=&y \left(\frac{2}{p_{\infty
   }^2}+4\right)+y^2 \left(\frac{2 \pi }{p_{\infty
   }^2}+3 \pi \right)+y^3 \left(\frac{24}{p_{\infty
   }^2}+\frac{4}{p_{\infty
   }^4}-\frac{2}{3 p_{\infty
   }^6}+\frac{80}{3}\right)+y^4 \left(\frac{30 \pi }{p_{\infty
   }^2}+\frac{9 \pi }{p_{\infty
   }^4}+\frac{105 \pi }{4}\right)\nonumber\\
   &+&y^5 \left(\frac{1120}{3 p_{\infty
   }^2}+\frac{160}{p_{\infty
   }^4}+\frac{16}{p_{\infty
   }^6}-\frac{4}{3 p_{\infty
   }^8}+\frac{2}{5 p_{\infty
   }^{10}}+\frac{1344}{5}\right)+y^6 \left(\frac{945 \pi }{2
   p_{\infty }^2}+\frac{525 \pi
   }{2 p_{\infty }^4}+\frac{50 \pi
   }{p_{\infty }^6}+\frac{1155 \pi
   }{4}\right)+\nonumber\\
   &+&y^7 \left(\frac{29568}{5 p_{\infty
   }^2}+\frac{4032}{p_{\infty
   }^4}+\frac{1120}{p_{\infty
   }^6}+\frac{80}{p_{\infty
   }^8}-\frac{24}{5 p_{\infty
   }^{10}}+\frac{4}{5 p_{\infty
   }^{12}}-\frac{2}{7 p_{\infty
   }^{14}}+\frac{109824}{35}\right
   )\nonumber\\
   &+&y^8 \left(\frac{15015 \pi }{2
   p_{\infty }^2}+\frac{24255 \pi
   }{4 p_{\infty }^4}+\frac{2205
   \pi }{p_{\infty }^6}+\frac{1225
   \pi }{4 p_{\infty
   }^8}+\frac{225225 \pi
   }{64}\right)\nonumber\\
   &+&y^9 \left(\frac{658944}{7
   p_{\infty }^2}+\frac{439296}{5
   p_{\infty
   }^4}+\frac{39424}{p_{\infty
   }^6}+\frac{8064}{p_{\infty
   }^8}+\frac{448}{p_{\infty
   }^{10}}-\frac{64}{3 p_{\infty
   }^{12}}+\frac{96}{35 p_{\infty
   }^{14}}-\frac{4}{7 p_{\infty
   }^{16}}+\frac{2}{9 p_{\infty
   }^{18}}+\frac{2489344}{63}\right)\nonumber\\
   &+&y^{10} \left(\frac{3828825 \pi
   }{32 p_{\infty
   }^2}+\frac{2027025 \pi }{16
   p_{\infty }^4}+\frac{135135 \pi
   }{2 p_{\infty }^6}+\frac{72765
   \pi }{4 p_{\infty
   }^8}+\frac{3969 \pi }{2
   p_{\infty }^{10}}+\frac{2909907
   \pi }{64}\right)+\dots
\eea
where we have defined
\beq
E=\gamma \mu_5,\quad J=M j \mu_5,\quad \gamma=\sqrt{1+p_\infty^2},\quad j=\frac{p_\infty}{y}\,.
\eeq
Note that the massless limit $O(p_\infty^0)$ of this expression coincides with the corresponding term in Eq. \eqref{chi_4d}, showing the special role of the massless case in both these geometries that at fixed $y$ are simply related by a Weyl rescaling.

Finally, Eq. \eqref{elliptic} admits an explicit representation in term of complete ($K$) or incomplete ($F$) elliptic integrals

\beq
\frac{\chi+\pi}{2}{=}\frac{\hat b}{2 f_1}\sqrt{1-f_1^2} \left[\sqrt{1-f_1^2} \left(\frac{\zeta}{2} - {\rm arcsinh}\left(\frac{f_1}{\sqrt{1{-}f_1^2}}\right)\right)
+  \sqrt{1-f_2^2} \left(K\left(\frac{f_2^2}{f_1^2}\right){+}i F\left(\frac{i
   \zeta }{2}|1-\frac{f_2^2}{f_1^2}\right)\right)\right]\,,
\eeq
where $\hat b=b/M$ and
\beq
\zeta=\log\left(\frac{1+f_1}{1-f_1}\right),\quad f_{1,2}=\sqrt{1-\frac{4M}{r_{1,2}}}\,,
\eeq
with $r_{1,2}/M$ defined in Eq. \eqref{r12sol} and depending only on $\hat E$ and $j$.
Equivalently, in terms of hypergeometric functions
\bea
\frac{\chi+\pi}{2}&=&\frac{\pi ^{3/2}}{2} \sum_{k=0}^\infty p_\infty^{-2 k} y^k \left(\frac{\,
   _3\tilde{F}_2\left(\frac{1}{4},\frac{3}{4},\frac{k}{2}+\frac{1}{2};\frac{1}{2},1-\frac{k}{2};16
   y^2\right)}{\Gamma \left(\frac{k}{2}+1\right)}+\frac{y \,
   _3\tilde{F}_2\left(\frac{3}{4},\frac{5}{4},\frac{k}{2}+1;\frac{3}{2},\frac{3}{2}-\frac{k}{2};16
   y^2\right)}{\Gamma \left(\frac{k+1}{2}\right)}\right)\,.
\eea
\end{widetext}

In the next section we will study the case of a minimally coupled, massless scalar field,  sourced by a scalar particle moving along a circular timelike geodesic of the 5d background, repeating in this case previous computations already performed in the past for Schwarzschild BHs and, more recently, for Top Stars  \cite{Bianchi:2024vmi,Bianchi:2024rod,DiRusso:2025lip,Bianchi:2025aei}.

\section{Scalar field in 5d}

A massless scalar field in the $W$-soliton spacetime, in the  minimal coupling case, evolves according the following wave equation
\beq
\label{box}
\Box \psi=-4\pi \rho_S\,,
\eeq
where
\beq
\label{rho_def}
\rho_S=q_S\int \frac{d\tau}{\sqrt{-g}} \delta^{(5)}(x-x_p(\tau))\,.
\eeq
is a scalar charge density, with $q_S$ a scalar `charge', describing the tri-linear coupling of a massive probe to the massless scalar field.
The latter is taken as a proxy for the more involved case of gravitational waves and may represent one of the two scalar fields sourcing the $W$-soliton. In particular setting  ${\rm Im}(z) = \exp(\sqrt{2}\psi/\sqrt{3})$ gives rise to a canonically normalized scalar field minimally coupled to the gravitational field.

Eq. \eqref{box} can be reduced by expanding in spherical harmonics and Fourier transforming the time
\beq
\label{psi_exp}
\psi(t,r,\theta,\phi,y)=\sum_{\ell mn} Y_{\ell m}(\theta,\phi)\int \frac{d\omega}{2\pi}e^{-i\omega t }e^{i \frac{n}{R_y} y}R_{\ell m \omega n}(r)\,.
\eeq
Notice that $R_{\ell m \omega n}(r)$ is actually $m$ independent in the homogeneous case, but it acquires a non-trivial dependence on $m$ in presence of a (generic) source.  
 
\subsection{Solutions of the homogeneous radial equation}
Taking $n=0=P_y$ for simplicity, the homogeneous radial equation is then given by
\bea
&& r(r-4M)R_{\ell\omega}''(r)+2(r-2M)R_{\ell\omega}'(r)\nonumber\\
&&+[r^2\omega^2-\ell(\ell+1)]R_{\ell\omega}(r)=0\,,
\eea
and admits two independent solutions, labeled below as \lq\lq in-solutions" (regular at $r=4M$) and \lq\lq up-solutions" (purely outgoing at $r\to +\infty$).
We will find these solutions in Post-Newtonian (PN) sense, determining 
\begin{enumerate}
\item  pure PN in- and up-solutions: $R_{\ell\omega}^{\rm in, PN}(r)$ and $R_{\ell\omega}^{\rm up, PN}(r)=R_{-\ell-1,\omega}^{\rm in, PN}(r)$, for generic $\ell$;
\item MST-type  in- and up-solutions:  $R_{\ell\omega}^{\rm in, MST}(r)$ and $R_{\ell\omega}^{\rm up, MST}(r)$, for specific values of $\ell=0,1,2,3,\ldots$
\end{enumerate}
They are discussed below where we also display the beginning of the power series representing them.

\subsection{PN-type solutions}
The in-solution is given by
\bea
R_{\ell\omega}^{\rm in, PN}(r)&=& r^\ell -\eta^2 
r^{\ell-1}  \frac{ 2 \ell M +r^3 \omega^2 }{2(3 + 2 \ell) }  \nonumber\\
&+&
\frac{\eta^4}{8}  r^{\ell-2} \left(\frac{r^6 \omega ^4}{(2\ell+3)(2\ell+5)}+\frac{32 (\ell-1)^2 l M^2}{2 \ell-1}\right.\nonumber\\
&+&\left. \frac{8 ((\ell-1) \ell-4) M r^3 \omega
   ^2}{(\ell+1) (2 \ell+3)}\right) +O(\eta^6)\,,
\eea
while the PN up-solution follows easily from the in-solution, via the relation mentioned above
$R_{\ell\omega}^{\rm up, PN}(r)=R_{-\ell-1,\omega}^{\rm in, PN}(r)$.

\subsection{MST-type solutions}

Following the MST prescriptions one has
\bea
R_{\ell\omega}^{\rm in, MST}(r)&=& \sum_{n=-\infty}^\infty a_n {}_2F_1(-n-\nu,1+n+\nu,1; z)\,,\qquad
\eea
with $\nu=\ell + ...$ denoting the `renormalized angular momentum' and 
\beq
z=-\frac{r-4M}{4M}\,,
\eeq
$R_{\ell\omega}^{\rm in, MST}$ can be expanded in PN sense e.g., $M\to M\eta^2$, $\omega \to \omega\eta $ with $\eta=1/{c}$ a placeholder for the PN expansion.
Similarly,
\bea
R_{\ell\omega}^{\rm up, MST}(r)&=& \sum_{n=-\infty}^\infty a_n e^{i \tilde z} \left(-2i\tilde z\right)^{1+n+\nu}\times \nonumber\\
&& U(1+n+\frac{i}{2}\epsilon,2 (1+n+\nu); -2i\tilde z) \,,\qquad
\eea
with $U(a,b;x)$ denoting Tricomi confluent hypergeometric function and 
\beq
\tilde z=\omega r\,,\qquad \epsilon=-4M\omega\,.
\eeq

The coefficients $a_n$ satisfy the following recursion relation
\beq
a_{n+1}\alpha_n+a_{n}\beta_n
+a_{n-1}\gamma_n=0
\eeq
with
\bea
\alpha_n &=& -\frac{i\epsilon (1+n+\nu)(1+n+\nu-\frac{i}{2}\epsilon)}{3+2n +2\nu}\,, \nonumber\\
\beta_n  &=& (\ell+\frac12)^2-(\nu+\frac12)^2 -n(n+1+2\nu)-\frac12 \epsilon^2\,,\nonumber\\
\gamma_n &=& -\frac{i\epsilon (n+\nu)(n+\nu+\frac{i}{2} \epsilon)}{-1+2n +2\nu}\,.
\eea

Let us recall that for the MST type solutions one identifies ($\ell$-by-$\ell$) the renormalized angular momentum variable $\nu$. We have checked that the $\nu$ obtained when working in MST coincides with the one obtained when working in the  qSW formalism, offering so far a consistency check that boundary conditions are the same in the two approaches (see also Refs. \cite{Bini:2025ltr,Bini:2025bll}).
For example, for $\ell=0$
\bea
\nu^{\ell=0}&=&0 -\frac{16 M^2 \omega ^2}{3}-\frac{3008 M^4 \omega ^4}{135}-\frac{11995136 M^6 \omega ^6}{42525}\nonumber\\ 
&-&\frac{1688869888 M^8 \omega ^8}{535815}+O((M\omega)^{10})\,.
\eea
For $\ell=1$
\bea
\nu^{\ell=1}&=&1-\frac{32 M^2 \omega ^2}{15}-\frac{101504 M^4 \omega ^4}{23625}-\frac{661221376 M^6 \omega ^6}{37209375}\nonumber\\
&-& \frac{12299100829696 M^8 \omega ^8}{128930484375}+O((M\omega)^{10})\,,
\eea
and
for $\ell=2$
\bea
\nu^{\ell=2}&=& 2-\frac{128
   M^2 \omega ^2}{105}-\frac{760064 M^4 \omega ^4}{1157625}\nonumber\\
&-&\frac{140668104704
   M^6 \omega ^6}{140390971875}-\frac{102919314892525568 M^8 \omega ^8}{44267379296765625}\nonumber\\
& +& O((M\omega)^{10})\,.
\eea

The in and up MST type solutions for $\ell=0,1,2$ then read 
\begin{widetext}
\bea
R_{0, \omega}^{\rm in, MST}(r)&=&1-\frac{1}{6} \eta ^2 r^2 \omega ^2+2 i \eta ^3 M \omega+\eta ^4 \left(\frac{r^4 \omega ^4}{120}-\frac{4}{3} M r \omega ^2\right)+O\left(\eta^5\right)\,,\nonumber\\
R_{0, \omega}^{\rm up, MST}(r)&=&\frac{1}{r \omega }+i \eta+\eta ^2 \left(\frac{2 M}{r^2 \omega }-\frac{r \omega }{2}\right)-\frac{i \eta ^3 \left(12 \gamma  M+r^3 \omega ^2\right)}{6 r}\nonumber\\
&+&\frac{\eta ^4 \left(\frac{128 M^2}{r^3}-96 M \omega ^2 \log (2 r \omega )+24
   (5-2 \gamma +2 i \pi ) M \omega ^2+r^3 \omega ^4\right)}{24 \omega }+O\left(\eta^5\right)\,,\nonumber\\
R_{1, \omega}^{\rm in, MST}(r)&=&\frac{r}{2 M}+\eta ^2 \left(-\frac{r^3 \omega ^2}{20 M}-1\right)+i \eta ^3 r \omega+\eta ^4 \left(\frac{r^5 \omega ^4}{560 M}-\frac{r^2 \omega ^2}{5}\right)+O\left(\eta^5\right)\,,\nonumber\\
R_{1, \omega}^{\rm up, MST}(r)&=&\frac{i}{r^2 \omega ^2}+\eta ^2 \left(\frac{4 i M}{r^3 \omega ^2}+\frac{i}{2}\right)+\eta ^3 \left(\frac{2 (\gamma -1) M}{r^2 \omega }-\frac{r \omega }{3}\right)+\frac{i \eta ^4 \left(\frac{576 M^2}{\omega ^2}+80 M r^3-5 r^6 \omega
   ^2\right)}{40 r^4}+O\left(\eta^5\right)\,,\nonumber\\
R_{2, \omega}^{\rm in, MST}(r)&=&\frac{3 r^2}{8 M^2}-\frac{3 \eta ^2 \left(56 M r+r^4 \omega ^2\right)}{112 M^2}+\frac{3 i \eta ^3 r^2 \omega }{4 M}+\eta ^4 \left(\frac{r^6 \omega ^4}{1344 M^2}-\frac{r^3 \omega ^2}{28
   M}+1\right)+O\left(\eta^5\right)\,,\nonumber\\
R_{2, \omega}^{\rm up, MST}(r)&=& -\frac{3}{r^3\omega^3}-\frac{\eta ^2 \left(36 M+r^3 \omega ^2\right)}{2 r^4 \omega ^3}+\frac{3 i (2 \gamma -3) \eta ^3 M}{r^3 \omega ^2}+\eta ^4 \left(-\frac{576 M^2}{7 r^5 \omega ^3}-\frac{4 M}{r^2 \omega }-\frac{r
   \omega }{8}\right)+O\left(\eta^5\right)\,.\nonumber\\
\eea
\end{widetext}

\subsection{The inhomogeneous radial equation}

As in Ref.~\cite{Bianchi:2024vmi}, we assume the source moving on a timelike circular equatorial geodesic, with parametric equations
\bea
t&=&t_p(\tau)=\Gamma \tau\,,\quad r=r_p(\tau)=r_0\,,\quad \theta=\theta_p(\tau)=\frac{\pi}{2}\,,\nonumber\\
\phi&=&\phi_p(\tau)=\Gamma \Omega \tau\,,\quad y_p(\tau)=0\,,\qquad
\eea
with $r_0>4M$ and 
\beq
\label{Omega_circ}
\Omega=\sqrt{\frac{M}{r_0^3-Mr_0^2}} \,,\qquad \Gamma =\frac{\sqrt{1-\frac{M}{r_0}}}{1-\frac{2M}{r_0}}\,,
\eeq
and 5-velocity 
\beq
U^\mu =\frac{dx^\mu_p(\tau)}{d\tau}=\Gamma (\delta^\mu_t+\Omega \delta^\mu_\phi)\,.
\eeq

The scalar charge density $\rho_S$, Eq. \eqref{rho_def}, can then be written as
\bea
\rho_S&=& q_S\int \frac{d\tau}{r(r-2M)\sin \theta}\delta^{(5)}(x-x_p(\tau)) \,,
\eea
where now $\sqrt{-g}=r(r-2M)\sin \theta=r^2 f_s(r)\sin \theta$ (notice the presence of the factor $f_s(r)$ in comparison with the familiar Schwarzschild spacetime) and
\bea
\label{delta5}
\delta^{(5)}(x-x_p(\tau))&=& \delta(t-\Gamma \tau)\delta(r-r_0)\delta(\theta-\frac{\pi}{2})\times \nonumber\\
&&\times \delta(\phi-\Omega t)\delta(y)\,.
\eea
The integral over $\tau$ is immediate and gives
\bea
\label{rho_def1}
\rho_S&=& \frac{q_S} {r_0^2f_s(r_0)\Gamma}\delta(r-r_0)\delta(\theta-\frac{\pi}{2}) \delta(\phi-\Omega t)\delta(y)\nonumber\\
&=& \frac{q_S} {r_0^2f_s(r_0)\Gamma} \sum_{\ell mn}\delta(r-r_0) Y_{\ell m}(\theta,\phi)Y_{\ell m}^*(\frac{\pi}{2},0)\times \nonumber\\
&\times & \int \frac{d\omega}{2\pi}e^{-i\omega t}\delta(\omega-m\Omega)\frac{e^{i n \frac{y}{R_y}}}{2\pi R_y}\,,
\eea
where we have used a Fourier series representation  for $\delta(y)$ 
\beq
\delta(y)=\sum_{n=-\infty}^\infty \frac{e^{i n \frac{y}{R_y}}}{2\pi R_y}\,,
\eeq
(instead of an integral representation) since $y$ is a periodic variable, 
besides the identity  
\bea
\delta(\theta-\frac{\pi}{2})\delta(\phi-\phi_0)
&=&\sum_{\ell m} Y_{\ell m}(\theta,\phi)Y_{\ell m}^*(\frac{\pi}{2},\phi_0)\,.\qquad
\eea
Consequently, the scalar charge density $\rho_S$ can be cast in the form
\bea
\label{rho_t2}
\rho_S 
&=& \sum_{\ell mn}{\mathcal S}_{\ell mn}(t,r)e^{i n \frac{y}{R_y}} Y_{\ell m}(\theta, \phi)\nonumber\\
&=&  \sum_{\ell mn}\int_{\omega}  e^{-i\omega t}\widehat {\mathcal S}_{\ell mn}(\omega,r)e^{i n \frac{y}{R_y}} Y_{\ell m}(\theta, \phi)
\,, 
\eea
where we have introduced the notation
\beq
\int_{\omega}=\int \frac{d\omega}{2\pi}\,.
\eeq
Furthermore, for dimensional reasons and to ease comparison with 4d spacetimes (i.e., the Schwarzschild spacetime), it is convenient to define
\beq
q_{\rm W}=\frac{q_S}{2\pi R_y}\,.
\eeq 
Therefore,
\bea
{\mathcal S}_{\ell mn}(t,r)&=& \int_{\omega}e^{-i\omega t}\widehat {\mathcal S}_{\ell mn}(\omega, r)\nonumber\\
&=&  \frac{q_{W}} {r_0^2f_s(r_0)\Gamma}\delta(r-r_0) e^{-im\Omega t}Y_{\ell m}^*(\frac{\pi}{2},0)\,,\nonumber\\
\widehat {\mathcal S}_{\ell mn}(\omega,r)&=&\int dt   e^{i\omega t}   {\mathcal S}_{\ell mn}(t,r)\nonumber\\
&=&   \frac{q_{W}} {r_0^2f_s(r_0)\Gamma} \delta(r-r_0)Y_{\ell m}^*(\frac{\pi}{2},0)\times \nonumber\\
&& 2\pi \delta(\omega-m\Omega)\,. 
\eea
In the present case $\widehat {\mathcal S}_{\ell mn}(\omega,r)$  does not depend explicitly on $n$. In fact one can `smear' the source along the $y$ direction or equivalently consider a string wrapped around the circle $S^1_y$ and focus on the $n=0$ component of the wave. The KK excitations with $n\neq 0$ are 'massive' and short-ranged. 

Let us then look for massless scalar waves $\psi$ in a Fourier series/integral expansion of the form \eqref{psi_exp}.
\begin{widetext}
Eq. \eqref{box}  implies
\bea
\Box \psi &=&\sum_{\ell mn} \int_{\omega} e^{-i \omega t}e^{in \frac{y}{R_y}}Y_{\ell m}(\theta,\phi)\left[\frac{f^2}{f_s}R_{\ell mn\omega}''+\frac{2}{r}R_{\ell mn\omega}' +\left(\frac{\omega^2}{f^2}- \frac{n^2 f_s^2}{R_y^2 f^2} -\frac{L}{r^2 f_s}\right)R_{\ell mn\omega}
\right]\nonumber\\
&=& -4\pi  \sum_{\ell m  n}\int_{\omega}  e^{-i\omega t}e^{in \frac{y}{R_y}}\widehat {\mathcal S}_{\ell mn}(\omega,r) Y_{\ell m}(\theta, \phi)\,,
\eea
(here $L=\ell(\ell+1)$) 
that is 
\beq
\sum_{\ell mn} Y_{\ell m}(\theta,\phi)\int_{\omega}e^{-i\omega t} e^{i n \frac{y}{R_y}}{\mathcal E}_{\ell mn\omega}(r)=0\,,
\eeq
where
\bea
{\mathcal E}_{\ell mn\omega}(r)&=& \frac{f^2}{f_s}R_{\ell mn\omega}''+\frac{2}{r}R_{\ell mn\omega}' +\left(\frac{\omega^2}{f^2}- \frac{n^2 f_s^2}{R_y^2 f^2} -\frac{L}{r^2 f_s}\right)R_{\ell mn\omega}
 +4\pi  \widehat {\mathcal S}_{\ell mn}(\omega,r) \,,\qquad
\eea
which will be solved  mode-by-mode by requiring ${\mathcal E}_{\ell mn\omega}(r)=0$.
Explicitly, we find
\bea
\label{eq_rad_R_finale}
\frac{f^2}{f_s}R_{\ell mn\omega}''+\frac{2}{r}R_{\ell mn\omega}' +\left(\frac{\omega^2}{f^2}- \frac{n^2 f_s^2}{R_y^2 f^2} -\frac{L}{r^2 f_s}\right)R_{\ell mn\omega} 
&=& -\frac{q_{W}} {r_0^2f_s(r_0)\Gamma} \delta(r-r_0)Y_{\ell m}^*(\frac{\pi}{2},0) 2\pi \delta(\omega-m\Omega)\,.
\eea
Eq. \eqref{eq_rad_R_finale} is conveniently rewritten by introducing the operator
\beq
\label{L_r_def0}
{\mathcal L}_{(r)}(\cdot )\equiv\frac{d^2}{dr^2}(\cdot )+\frac{2f_s}{rf^2}\frac{d}{dr}(\cdot )+\frac{f_s}{f^2}\left(\frac{\omega^2}{f_s(r)}- \frac{n^2}{R_y^2 f_b(r)}-\frac{L}{r^2}\right)(\cdot ) \,,
\eeq
\end{widetext}
as well as the combination
\bea
S_{\ell mn\omega}&=&- \frac{4\pi q_{\rm W}} {\Gamma \Delta(r_0)} Y_{\ell m}^*(\frac{\pi}{2},0) 2\pi \delta(\omega-m\Omega)\nonumber\\
&=&\bar S_{\ell mn\omega}2\pi \delta(\omega-m\Omega)\,,
\eea
with
\bea
\bar S_{\ell mn\omega}  
&=&-q_{\rm W}K(r_0)Y_{\ell m}^*(\frac{\pi}{2},0)\,,
\eea
where we have also defined
\beq
K(r_0)= \frac{4\pi }{\Gamma \Delta(r_0)}\,,\qquad \Delta=r^2 f^2\,.
\eeq
Eq. \eqref{eq_rad_R_finale} then becomes
\bea
{\mathcal L}_{(r)}R_{\ell mn\omega}&=& S_{\ell mn\omega} \delta(r-r_0)\,.
\eea
We focus here of the \lq\lq smearing limit" $n=0$ mode, namely we limit our considerations to the lowest KK mode in order to facilitate comparisons with 4d spacetimes.

This equation can be solved by using the Green's function method, {\it i.e.} introducing the Green's function  
\beq
{\mathcal L}_{(r)} G_{\ell\omega}(r,r')=\frac{1}{\Delta(r')}\delta(r-r')\,, 
\eeq
with
\bea
G_{\ell\omega}(r,r')&=&\frac{1}{W_{\ell\omega}}[R_{\rm in}(r)R_{\rm up}(r')H(r'-r)\nonumber\\
&+&R_{\rm in}(r')R_{\rm up}(r)H(r-r')]\nonumber\\
&\equiv & \frac{1}{W_{\ell\omega}}R_{\rm in}(r_<)R_{\rm up}(r_>)\,,
\eea
where $H(x)$ the Heaviside step function, $r_<$ and $r_>$ corresponding to $r$, $r'$,  and $R_{\rm in}(r)$ and $R_{\rm up}(r)$ are two independent solutions of the homogeneous equation, both depending on $\ell$ and $\omega$ but neither on $m$ thanks to spherical symmetry nor on $n$ since we consider only massless scalar waves with $n=0$. They can be written in various forms, e.g. PN, MST.  

The physical range of the radial variable is $[4M,\infty)$ and
\beq
W_{\ell\omega}=\Delta (r) [R_{\rm in}(r)R_{\rm up}'(r)-R_{\rm up}(r)R_{\rm in}'(r)]\,,
\eeq
is the constant Wronskian.
Finally,
\bea
\label{final_R}
R_{\ell mn\omega}(r)&=&S_{\ell mn\omega} \int dr' G_{\ell\omega}(r,r')\Delta(r')  \delta(r'-r_0)\nonumber\\
&=&S_{\ell mn\omega}G_{\ell\omega}(r,r_0) \,.
\eea

With this expression for the radial functions  we can evaluate the field
\begin{widetext}
\bea
\psi(t,r,\theta,\phi,y)&=&\sum_{\ell mn}\int_{\omega }e^{-i\omega t}e^{in\frac{y}{R_y}}R_{\ell mn\omega}(r)Y_{\ell m}(\theta,\phi)\nonumber\\
&=& \sum_{\ell mn}\int_{\omega }e^{-i\omega t}e^{in\frac{y}{R_y}}\bar S_{\ell mn\omega}G_{\ell\omega}(r,r_0)2\pi \delta(\omega-m\Omega) Y_{\ell m}(\theta,\phi)\nonumber\\
&=& \sum_{\ell mn} e^{in\frac{y}{R_y}}\bar S_{\ell mn\omega}G_{\ell\omega}(r,r_0)e^{-im\Omega t} Y_{\ell m}(\theta,\phi)\nonumber\\
&=& -q_{\rm W} K(r_0) \sum_{\ell mn} e^{in\frac{y}{R_y}} G_{\ell\omega}(r,r_0) Y_{\ell m}(\theta,\phi-\Omega t) Y_{\ell m}^*(\frac{\pi}{2},0)\,, 
\eea
which along the orbit becomes
\beq
\psi(t,r_0,\frac{\pi}{2},\Omega t,0)
= -q_{\rm W} K(r_0) \sum_{\ell mn}|Y_{\ell m}(\frac{\pi}{2},0) |^2   G_{lmn\omega}(r_0,r_0)   \,,
\eeq
{\it i.e.}   it does not depend on $t$. 
\end{widetext}

\subsection{Energy loss along circular orbits}

Let us consider the loss of energy and angular momentum by emission of (massless) scalar waves  
which follow  from the associated energy-momentum tensor.   
For a canonically normalized massless complex scalar field it is given by
\beq
8\pi T^{\rm scal}_{\mu\nu}=\partial_\mu\psi^*\partial_\nu\psi+ \partial_\mu\psi \partial_\nu\psi^* -  g_{\mu\nu} \partial_\lambda \psi^* \partial^\lambda\psi \,,
\eeq
so that
\beq
\frac{d^2E}{dtd\Omega}=\lim_{r\to \infty} (r^2 T^{\rm scal}{}^r{}_t)\,.
\eeq
We find
\beq
8\pi T^{\rm scal}_{rt}=\psi^*_{,r}\psi_{,t}+\psi_{,r}\psi^*_{,t} \,,
\eeq
implying
\bea
8\pi T^{\rm scal}{}^r{}_{t}&=& \frac{f^2}{f_s} (\psi^*_{,r}\psi_{,t}+\psi_{,r}\psi^*_{,t})\nonumber\\
&=& \frac{f^2}{f_s} \psi^*_{,r}\psi_{,t}+{\rm c.c.}\,.
\eea

Here, suppressing the dependence on the spacetime variables to ease notation and assuming only emission of massless $n=0$ KK modes, without loss of generality since, we recall,  higher KK would be massive from a 4d perspective,
\bea
\psi&=& \sum_{\ell m}Y_{\ell m}\int \frac{d\omega}{2\pi}e^{-i\omega t}R_{\ell m\omega}\,. 
\eea 
Therefore
\begin{widetext}
\bea
r^2 T^{\rm scal}{}^r{}_{t}&=&\frac{r^2  f^2 }{8\pi f_s} (\psi^*_{,r}\psi_{,t}+\psi_{,r}\psi^*_{,t})\nonumber\\
&=& \frac{r^2f^2  }{8\pi f_s} \sum_{\ell m,\ell 'm'}\int \frac{d\omega}{2\pi} \frac{d\omega'}{2\pi}Y_{\ell m}(\theta,\phi)Y_{\ell 'm'}^*(\theta,\phi)  \left[(-i\omega) e^{-i(\omega-\omega')t} R_{\ell m\omega}(r)\frac{d}{dr}R^*_{\ell 'm'\omega'}(r)\right.\nonumber\\
&+& \left. (i\omega')e^{i(\omega-\omega')t} R^*_{\ell 'm'\omega'}(r)\frac{d}{dr}R_{\ell m \omega }(r)\right]\,.
\eea
\end{widetext}

\hbox{\vspace{2cm}}

Integrating over the 2-sphere and recalling the orthogonality property  
\beq
\int \sin \theta d\theta d\phi   Y_{\ell m}^*(\theta,\phi)Y_{\ell 'm'}(\theta,\phi)=\delta_{\ell\ell'}\delta_{m m'}\,,
\eeq
we find
\bea
\frac{dE}{dt}&=&\lim_{r\to \infty}\int \sin \theta d\theta d\phi r^2 T^{\rm scal}{}^r{}_{t}\nonumber\\
&=&\lim_{r\to \infty} \frac{\Delta (r)}{8\pi} \sum_{\ell m}\int \frac{d\omega}{2\pi} \frac{d\omega'}{2\pi}\Big[ \nonumber\\
&&(-i\omega) e^{-i(\omega-\omega')t} R_{\ell m\omega}(r)\frac{d}{dr}R^*_{\ell m\omega'}(r)\nonumber\\
&+&  (i\omega')e^{i(\omega-\omega')t} R^*_{\ell m\omega'}(r)\frac{d}{dr}R_{\ell m \omega }(r)\Big]\,,
\eea
with  $\Delta (r)=r^2 f^2$.
Because of the $r\to \infty$ limit we can use here only the up-part of the $R_{\ell m\omega}(r)$ solution, namely
\bea
\label{sol_con_Z}
R_{\ell m\omega}(r)
&=& \frac{R_{\rm up}(r)}{W_{lm\omega}}  R_{\rm in}(r_0) \Delta (r_0)\bar S_{\ell m\omega}2\pi \delta(\omega-m\Omega)\nonumber\\
&=& {\mathfrak R}_{\ell m\omega}(r)2\pi \delta(\omega-m\Omega)
\,,
\eea
with
\bea
 {\mathfrak R}_{\ell m\omega}(r)&=&- \frac{4\pi q_{\rm W}} {\Gamma}\frac{R_{\rm in}(r_0)  Y_{\ell m}^*(\frac{\pi}{2},0)}{W_{\ell m\omega}}  R_{\rm up}(r)\nonumber\\
&=&q_{{\rm W}} C_{\ell m\omega}(r_0)R_{\rm up}(r)\,,
\eea
with
\beq
 C_{\ell m\omega}(r_0)=- \frac{4\pi} {\Gamma}\frac{R_{\rm in}(r_0)  Y_{\ell m}^*(\frac{\pi}{2},0)}{W_{\ell m\omega}}\,.
\eeq

Finally,
\begin{widetext}
\bea
\frac{dE}{dt}
&=&\lim_{r\to \infty} \frac{\Delta(r)}{8\pi} \sum_{\ell m}\int \frac{d\omega}{2\pi} \frac{d\omega'}{2\pi}\left[(-i\omega) e^{-i(\omega-\omega')t} R_{\ell m\omega}(r)\frac{d}{dr}R^*_{\ell m\omega'}(r)\right.\nonumber\\
&+& \left. (i\omega')e^{i(\omega-\omega')t} R^*_{\ell m\omega'}(r)\frac{d}{dr}R_{\ell m \omega }(r)\right]\nonumber\\
&=& \lim_{r\to \infty} \frac{\Delta(r)}{8\pi} \sum_{\ell m} (-im\Omega) \left[ {\mathfrak R}_{\ell m\omega}(r)\frac{d}{dr}{\mathfrak R}^*_{\ell m\omega}(r) -  {\mathfrak R}^*_{\ell m\omega}(r)\frac{d}{dr}{\mathfrak R}_{\ell m\omega}(r)\right]_{\omega=m\Omega}\nonumber\\
&=& \lim_{r\to \infty} \frac{\Delta(r)}{4\pi} \sum_{\ell m} m\Omega\,\,  {\mathcal  Im}\left({\mathfrak R}_{\ell m\omega}(r)\frac{d}{dr}{\mathfrak R}^*_{\ell m\omega}(r)\right)\big|_{\omega=m\Omega}\,.
\eea
\end{widetext}
Let us introduce the constants (i.e., not depending on $r$)
\bea
{\mathfrak F }_{\ell m}(r_0)&=& \lim_{r\to \infty} \Delta(r) {\mathcal  Im}\left({\mathfrak R}_{\ell m\omega}(r)\frac{d}{dr}{\mathfrak R}^*_{\ell m\omega}(r)\right)\big|_{\omega=m\Omega}\nonumber\\
F^{\rm up}_{\ell m}(r_0)&=& \lim_{r\to \infty} \Delta(r) {\mathcal  Im}\left(R_{\rm up}(r)\frac{d}{dr}R^*_{\rm up}\right)\big|_{\omega=m\Omega}\,,
\eea
and the combination
\beq
{\mathcal F }_{\ell m}(r_0)=
q_{{\rm W}}^2 |C_{\ell m\omega}(r_0)|^2\big|_{\omega=m\Omega}F^{\rm up}_{\ell m}(r_0)\,.
\eeq
Note that one cannot use the PN solutions for $R_{\rm in/up}(r)$ because these solutions are real, implying necessarily ${\mathcal F }_{\ell m}(r_0)=0$. In other words,
${\mathcal F }_{lm}(r_0)$ requires the MST solutions to be computed, and hence depends crucially from having imposed the correct boundary conditions for the solution of the radial equation.
 
Consequently,
\bea
\frac{dE}{dt}&=& \frac{q_{{\rm W}}^2}{4\pi}\sum_{\ell m} m\Omega |C_{\ell m\omega}(r_0)|^2\big|_{\omega=m\Omega}F^{\rm up}_{\ell m}(r_0)\,,
\eea
with
\bea
 |C_{\ell m\omega}(r_0)|^2 &=& \frac{16\pi^2} {\Gamma^2}\frac{|R_{\rm in}(r_0)|^2  |Y_{\ell m}(\frac{\pi}{2},0)|^2}{|W_{\ell m\omega}|^2}\nonumber\\
&=&  16\pi^2  \left(1-\frac{3M}{r_0}\eta^2\right) \times \nonumber\\
&& \frac{|R_{\rm in}(r_0)|^2  |Y_{\ell m}(\frac{\pi}{2},0)|^2}{|W_{\ell m\omega}|^2}\,.
\eea
Separating the contributions of the various spherical harmonics one gets
\bea
\frac{dE}{dt}&=&\sum_{\ell =0}^\infty \left(\frac{dE}{dt}\right)^{(\ell)}\,,  
\eea
In the present investigation we limit our considerations to the $\ell=0,1,2$ modes.   
For instance
\beq
\left(\frac{dE}{dt}\right)^{\ell=0}=\frac{q_{{\rm W}}^2}{4\pi} \sqrt{\frac{r_s}{2 r_0^{3}}}\lim_{m\to 0}\left(  m |C_{0m\omega}(r_0)|^2\big|_{\omega=m\Omega}F^{\rm up}_{0m}(r_0)\right)\,.
\eeq
Recalling the definition  $u_0={M}/{r_0}$, 
and limiting our considerations at the NNLO approximation level we find (omitting the overall factor $q_S^2/(4\pi^2 R_y^2 M^2)=q_{{\rm W}}^2/M^2$)\\
\bea
\left(\frac{dE}{dt}\right)^{\ell=0}&=& O(\eta^{10})\,,\nonumber\\
\left(\frac{dE}{dt}\right)^{\ell=1}&=& \frac{\pi q_W^2}{M^2}u_0^2 \Big[-\frac{8}{9} \eta ^3 u_0^2+\frac{208 \eta ^5 u_0^3}{45}-\frac{7864 \eta ^7 u_0^4}{1575}\nonumber\\
&+&\frac{8 \eta ^9 u_0^5 \left(10080 \log (4 u e^{2\gamma}){-}6300 \pi ^2 
   {-}120317\right)}{42525}\Big]\nonumber\\
&+& O(\eta^{10})\,,\nonumber\\
\left(\frac{dE}{dt}\right)^{\ell=2}&=& \frac{\pi q_W^2}{M^2}u_0^2  \left[{-}\frac{104}{45}u_0^3\eta^5{+}\frac{3464}{175}u_0^4\eta^7{-}\frac{187288}{3969}u_0^5\eta^9\right]\nonumber\\
&+& O(\eta^{10})\,.\nonumber\\
\eea
The sum of the various contributions yields
\bea
    \frac{dE}{dt}
&=& -\frac{8\pi q_W^2u_0^4}{9M^2}\Big[1-\frac{13}{5}u_0-\frac{2914u_0^2}{175}\nonumber\\
&+&u_0^3\left(-\frac{32}{15} \log (4 u_0 e^{2\gamma})+\frac{4 \pi ^2}{3}+\frac{2598044}{33075}\right)\Big]\,,\nonumber\\
   &+&O\left(u_0^6\right)
\eea
where $q_{{\rm W}} =\frac{q_S}{2\pi R_y}$,
and  
we put  $\eta=1$.  Note that, differently from the 4d case, where $q_4\sim L$ has the dimensions of a length, in the present five dimensional case $q_S\sim L^2$,  i.e.  it scales with $(2M)^2$ and not with $(2M)$.

Moreover, because both $t$ and $\phi$ are  coordinates adapted to the Killing vectors of the background, $\xi_{(t)}=\partial_t$ and $ \xi_{(\phi)}=\partial_\phi$, the angular momentum variation along the symmetry axis, $dL$, is simply related to the energy variation $dE$ by (see Eq. (4.13) of Ref. \cite{Teukolsky:1974yv} which also contains the relation between the loss of $L$ and the energy momentum tensor, $dE=T^{\mu\nu} \xi_{(t)\mu}d\Sigma_\nu$, $dL=T^{\mu\nu} \xi_{(\phi)\mu}d\Sigma_\nu$, etc.)
\beq
dL=\frac{m}{\omega}dE\,,
\eeq
where $m$ is the azimuthal quantum number.

\subsection{Final result}
The expression of the energy loss in terms of $u_0=M/r_0$ is given by
\bea
\label{DEDT_final}
\frac{dE}{dt}&=& -\frac{8\pi u_0^4}{9}\left(\frac{q}{M}\right)^2\left[  1 - \frac{13}{5} u_0 - \frac{2914}{175} u_0^2 \right.\nonumber\\  
&+&\left.  
 u_0^3 \left(\frac{2598044}{33075}    +  \frac{4}{3} \pi^2 - \frac{32}{15} \ln(4 e^{2\gamma_E} u)\right) \right]\,,
\eea
where, in the computation, we used only the multipoles $\ell=0,1,2$.
Since $u$ is an inverse radial variable it has a coordinate origin and hence it is a gauge dependent quantity.
Differently $M\Omega$ (the dimensionless angular velocity variable) is a gauge-invariant quantity, i.e., a physical observable.
Re-expressing the final result as a function of the gauge-invariant variable $x=(M\Omega)^{2/3}$, 
\bea
&&x=(M\Omega)^{2/3}=\frac{u_0}{(1-u_0)^{1/3}}\nonumber\\
&&=u_0 + \frac{u_0^2}{3} + \frac{2}{9} u_0^3  + \frac{14}{81} u_0^4  + \frac{35}{243} u_0^5+O(u_0^6) \, , \quad
\eea
with inverse
\beq
u_0=x  - \frac13 x^2  + \frac{1}{81}x^4  + \frac{1}{243}x^5 +O(x^7) \,,\qquad
\eeq
one finds
\bea
\frac{dE}{dt}&=& -\frac{8\pi  q^2 x^4}{9 M^2}\left[1 - \frac{59}{15} x  - \frac{2039}{175} x^2 \right.\nonumber\\ 
&+& \left.
 x^3 \left(\frac{10802158}{99225} + \frac{4}{3}\pi^2   -  \frac{32}{15}\ln (4e^{2\gamma_E}x) \right)  \right]\nonumber\\
&+& O(x^8)\,.
\eea

\subsection{Comparison}
Let us compare the results just obtained for the case of the $W$-soliton metric with the corresponding ones for Schwarzschild BHs and for Top Stars.
All the expressions take the form
\bea
\frac{dE_X}{dt}&=&-\left(\frac{q_X}{M}\right)^2 C_X u_0^4 {\mathcal E}_X(u_0)\,.
\eea
where $C_X$ denotes a \lq\lq form factor,"  different case by case.

In the Schwarzschild case \cite{Bini:2016egn}, namely $X={\rm Schw}$
\bea
C_{\rm Schw}&=& \frac13 \,,\nonumber\\
{\mathcal E}_{\rm Schw}(u_0)&=& 1-2u_0+2\pi u_0^{3/2}-10u_0^2\nonumber\\
&+& \frac{12}{5}\pi u_0^{5/2}+O(u_0^3)\,. 
\eea
Similarly, in the case of a topological star  \cite{DiRusso:2025lip}  
\bea
C_{\rm TS}&=& \frac13 \,,\nonumber\\
{\mathcal E}_{\rm TS}(u_0)&=& {\mathcal E}_0(u_0)+\alpha{\mathcal E}_1(u_0)+
\alpha^2 {\mathcal E}_2(u_0) \,,\qquad\nonumber\\
\eea
where
\bea
{\mathcal E}_0(u_0)&=& 1-2u_0-\frac{3117}{175}u_0^2+\frac{121984}{2205}u_0^3 +O(u_0^4)\,,  \nonumber\\
\nonumber\\
{\mathcal E}_1(u_0)&=& -2 u_0-5u_0^2+\frac{8768}{105}u_0^3 +O(u_0^4)\,, \nonumber\\
\nonumber\\
{\mathcal E}_2(u_0)&=& u_0^2+\frac{256}{15}u_0^3 +O(u_0^4)\,,\nonumber\\
\eea
with $\alpha=r_b/r_s$ an additional parameter entering the TS solution. Here ${\mathcal E}_k(u_0)$ starts with $u_0^k$, $k=0,1,2$ and $q_{_{\rm TS}} =\frac{q}{2\pi R_y}$.
In both these cases (Schwarzschild and TS)  $\Omega$ satisfies the Kepler law exactly
\beq
\Omega=\sqrt{\frac{M}{r_0^3}}\,,\qquad x=(M\Omega)^{2/3}=\frac{M}{r_0}=u_0\,,
\eeq
and hence $u_0=x$.
The above expressions can then be compared with the one obtained above for the $W$-soliton, Eq. \eqref{DEDT_final},
\bea
\frac{dE_{\rm W}}{dt}&=&-\left(\frac{q_{\rm W}}{M}\right)^2 C_{\rm W} x^4 {\mathcal E}_{\rm W}(x)\,,
\eea
namely,
\bea
C_{\rm W}&=& \frac89 \,,\nonumber\\
{\mathcal E}_{\rm W}(x)&=& 1 - \frac{59}{15} x  - \frac{2039}{175} x^2  \nonumber\\ 
&+& \left.
 x^3 \left(\frac{10802158}{99225} + \frac{4}{3}\pi^2   -  \frac{32}{15}\ln (4e^{2\gamma_E}x) \right)  \right]\nonumber\\
&+& O(x^8)\,.
\eea
As a general consideration, we see that all the expressions for energy loss all start with the same power $x^4$, with the numerical coefficient which is $\frac13$ either for Schwarzschild BH and TS, but $\frac89$ for the $W$-soliton. The expression for Schwarzschild BHs is the only one involving fractional powers of $x$, since this is the only solution with a horizon: in fact both the TS and the $W$-soliton contain a regular cap. Deviations from one to the other situations are enhanced as soon as one enters the strong field region, as expected.

\section{Concluding remarks}

We have studied the so-called neutral $W$-soliton solution, recently discovered in \cite{Chakraborty:2025ger, Dima:2025tjz}, which is an exact solution of the Einstein-Maxwell equations in 5 dimensions  with source two electromagnetic fields and, after reduction to 4 dimensions, a complex scalar field.
The interesting feature is that its reduced version in d=4 is also an exact solution which contains a naked singularity and its source is a (complex) scalar field.

In both cases, 4d and 5d, we have studied a scattering process of massless and massive particle on the background, reconstructing the scattering angle either with exact expressions or with large-angular momentum expansion expressions (which we have shown how to resum in some useful form).
Finally we have analized the case of a scalar field evolving on the background, characterizing Quasi-Normal-Modes in case of (non-)minimal coupling and the radiated energy in the case of minimal coupling. Our result for the energy loss is fully analytic and presented in PN expanded form, following the approach termed gravitational self force, although we focussed on massless scalar waves as a proxy of gravitational waves. This result has been obtained by  generalizing to this context the procedure developed by Mano-Suzuki-Takasugi for black holes and checking its consistency with the modern approach based on qSW and CFT.
This nontrivial accomplishment has allowed us to  compare and contrast with similar situations in the case of a Schwarzschild black hole and that of a topological star.

It will be a challenge left to future works to extend these type of studies  to the case of a charged $W$-soliton solution and to genuine GWs.

\section*{Acknowledgments}

We thank  I.~Bena, A.~Cipriani, A.~Dima, T.~Damour, F.~Fucito, A.~Geralico, P.~Heidmann, M.~Melis, J.~F.~Morales, P.~Pani, J.~Parra-Martinez, A.~Tokareva  for useful discussions. D.~B.  acknowledges sponsorship of
the Italian Gruppo Nazionale per la Fisica Matematica (GNFM) of the Istituto Nazionale di Alta Matematica
(INDAM). M.~B. and G.~D.~R. thank the MIUR PRIN contract 2020KR4KN2 ``String Theory as a bridge between Gauge Theories and Quantum Gravity'' and the
INFN project ST\&FI String Theory and Fundamental Interactions for partial support.

\appendix

\section{Additional geometrical characterization of the $W$-soliton metric}

In this section we will list various additional information aiming at a better characterization of  the $W$-soliton metric. In particular, the Cotton current and the Maxwell-like form of the Bianchi identities.

\subsection{Cotton tensor}
The (4-dimensional) \lq\lq Cotton tensor" (see e.g. Refs. \cite{Bini:2001ke,Bini:2004qf}) is defined as
\beq
J^{\alpha}{}_{\beta\gamma}=   \nabla_{[\beta}\left(R^\alpha{}_{\gamma]}-\frac16 R \delta^\alpha{}_{\gamma]}\right)\,,
\eeq
(the definition is modulo an overall sign, depending on conventions) and it is antisymmetric with respect the last two indices. It is used to re-express the divergence of the Weyl tensor as a consequence of the Bianchi identities of the second type, namely
\beq
\label{div_weyl}
\nabla_{\delta} C^{\alpha\delta}{}_{\beta\gamma}=J^{\alpha}{}_{\beta\gamma}\,,
\eeq
i.e., exactly as for Maxwell's equations in the presence of currents source of the electromagnetic field
\beq
\nabla_{\beta}F^{\alpha\beta}=4\pi J^\alpha\,.
\eeq
The only nonvanishing coordinate components of the Cotton tensor are 
\bea
J_{\phi\phi r } &=& 
\frac{M^2}{r^3f^2} \sin^2(\theta)\,, \nonumber\\
J_{\theta  r \theta  } &=& 
-\frac{M^2}{r^3f^2}\,, \nonumber\\
J_{trt } &=&  
-\frac{2M^2}{r^5f^2}\,, 
\eea
besides the obvious symmetry relations $J_{\phi\phi r }=-J_{\phi r \phi}$, $J_{\theta  r \theta  }= -J_{\theta\theta r}$, $J_{trt }= -J_{ttr}$.
Because of the antisymmetry in the last two indices one naturally introduces the spacetime dual of this tensor
\beq
J^{*}_{\alpha\beta\gamma}=\frac12 J_{\alpha\mu\nu}\eta^{\mu\nu}{}_{\beta\gamma}\,,
\eeq
also antisymmetric with respect to the last two indices\footnote{We denote by $\eta_{\mu\nu\rho\sigma} = \sqrt{|g|} \varepsilon_{\mu\nu\rho\sigma}$ the totally anti-symmetric Levi-Civita tensor.}, 
with nonvanishing coordinate components
\bea
J^{*}_{t\phi\theta} &=& 
-\frac{M^2}{r^3 f} \sin \theta  \,,\nonumber\\ 
J^{*}_{\theta\phi t} &=& 
-\frac{2M^2}{r^3 f}  \sin \theta  \,,\nonumber\\ 
J^{*}_{\phi \theta t} &=& 
\frac{2M^2}{r^3 f}\,. 
\eea
Both these tensors are used to form the Cotton complex combination
\beq
J^{(c)}_{\alpha\beta\gamma}=J_{\alpha\beta\gamma}+i J^{*}_{\alpha\beta\gamma}\,.
\eeq
With respect to the natural frame adapted to the fiducial observers $U=e_{\hat t}$, Eqs. \eqref{U_obs} and \eqref{spat_frame}, 
the nonvanishing frame components (denoted by the hat) of $J^{(c)}_{\alpha\beta\gamma}$ are much simplified
\bea
J^{(c)}_{\hat \phi \hat \phi \hat r}&=& J^{(c)}_{\hat \theta \hat \theta \hat r}=\frac{M^2}{r^5 f^{5/2}}\,, \nonumber\\
J^{(c)}_{\hat t \hat \phi \hat \theta}&=& J^{(c)}_{\hat \phi \hat t \hat \theta}=-i \frac{M^2}{r^5 f^{5/2}}\,, \nonumber\\
J^{(c)}_{\hat \theta \hat \phi \hat t}&=&-J^{(c)}_{\hat \phi \hat \theta \hat t}=    -2i \frac{M^2}{r^5 f^{5/2}},\nonumber\\ 
J^{(c)}_{\hat t \hat r \hat t}&=& -2\frac{M^2}{r^5 f^{5/2}}\,, 
\eea
modulo symmetries.
Ref. \cite{Bini:2004qf} has shown that $J^{(c)}_{\alpha\beta\gamma}$ is fully summarized by a spatial vector ($\rho^{\rm (G)}$) and a symmetric spatial 2-tensor
($J^{\rm (G)}_{\hat a \hat b}$) when aiming at rewriting Eq. \eqref{div_weyl} in a Maxwell-like form
\bea
\rho^{\rm (G)}_{\hat a}=-J^{(c)}_{\hat t \hat a \hat t}\,,\qquad J^{\rm (G)}_{\hat a \hat b}=-J^{(c)}_{(\hat a \hat b) \hat t}\,,
\eea
expressed here in components to avoid the introduction of the projector $P(U)=g+U\otimes U$ otherwise needed to identify the spatial parts.
We find 
\bea
\rho^{\rm (G)}&=& \frac{2M^2}{r^5 f^{5/2}}  e_{\hat r}\,,\qquad
J^{\rm (G)}= 0\,.
\eea

\subsection{Electric and magnetic parts of the Weyl tensor}

Let us  recall the definition of the electric and magnetic parts\footnote{Usually the definition of the magnetic part is modulo an overall sign.} of the Weyl tensor with respect to our fiducial observers $U$
\bea
E(U)_{\alpha\beta}&=&C_{\alpha \mu \beta \nu}U^\mu U^\nu\,,\nonumber\\
H(U)_{\alpha\beta}&=&-C^*_{\alpha \mu \beta \nu}U^\mu U^\nu\,,
\eea
summarized by the complex spatial 2-tensor $Z(U)=E(U)-i H(U)$.
In this case $H(U)_{\alpha\beta}=0$ and
\beq
E(U)=\frac{M}{fr^3}[-2 e_{\hat r}\otimes e_{\hat r}+e_{\hat \theta}\otimes e_{\hat \theta}+e_{\hat \phi}\otimes e_{\hat \phi}]\,,
\eeq
so that $Z(U)$ is real.

\subsection{Simon-Mars tensor}

Let us compute the kinematical fields of the observers: acceleration $a(U)$, given in Eq. \eqref{acc_U} above, and vorticity $\omega(U)$ (identically vanishing), summarized in the complex combination
\beq
z(U)=-a(U)-i\omega(U)=-\frac{M}{r^2 f^{3/2}}  e_{\hat r}\,,
\eeq
that is also real.
Furthermore
\beq
[z(U)\otimes z(U)]^{\rm TF}= \frac{M^2}{r^4 f^{3}}[e_{\hat r}\otimes e_{\hat r}]^{\rm TF}\,,
\eeq
where the trace-free (TF) projection yields
\bea
[e_{\hat r}\otimes e_{\hat r}]^{\rm TF}&=& e_{\hat r}\otimes e_{\hat r}-\frac{1}{3}[e_{\hat r}\otimes e_{\hat r}+e_{\hat \theta}\otimes e_{\hat \theta}+e_{\hat \phi}\otimes e_{\hat \phi}]\nonumber\\
&=& \frac{2}{3}e_{\hat r}\otimes e_{\hat r}-\frac{1}{3}[e_{\hat \theta}\otimes e_{\hat \theta}+e_{\hat \phi}\otimes e_{\hat \phi}]\nonumber\\
&=& -\frac13 [-2 e_{\hat r}\otimes e_{\hat r}+e_{\hat \theta}\otimes e_{\hat \theta}+e_{\hat \phi}\otimes e_{\hat \phi}]\nonumber\\
&=& -\frac13 \frac{fr^3}{M}E(U)
\eea
and then $[z(U)\otimes z(U)]^{\rm TF}$ and $E(U)$ are parallel, namely
\bea
[z(U)\otimes z(U)]^{\rm TF}= -\frac13 \frac{M}{r  f^{2}}E(U)\,.
\eea
Consequently, defining the Simon-Mars tensor \cite{simon,mars1999} (see also Refs. \cite{Bini:2001ke,Bini:2004qf})
\beq
S(U)=z(U)\times Z(U)\,,
\eeq
one immediately finds that 
\beq
S(U)=a(U)\times E(U)=0\,.
\eeq

\subsection{Maxwell-like equations}
Let us introduce the (1+3) split  versions of both divergence and (symmetric) curl operations for spatial 2-tensors as well as the notion of vector product of a vector with a 2 tensor.
Denoting by $x$ a generic spatial vector and by $X$ a generic spatial 2-tensor (with respect to $U$, i.e. orthogonal to $U$ with respect to both indices; notice that the request of being spatial tensors for $x$ and $X$ is crucial here) one defines
\bea
[{\rm div}(U)X]^\alpha&=& P(U)_{\mu}{}^{\nu}\nabla_\nu X^{\mu\alpha}\equiv \nabla(U)_\mu  X^{\mu\alpha}\,, \nonumber\\
{}[{\rm Scurl}(U)X]^{\alpha\beta}&=& \eta(U)^{\gamma \delta (\alpha}\nabla(U)_\gamma X^{\beta)}{}_\delta\,,\nonumber\\
{}[x \times X]^{\alpha\beta}&=& \eta(U)^{\gamma\delta (\alpha}x_\gamma X^{\beta)}{}_\delta\,, 
\eea
with $\eta(U)^{\alpha\beta\gamma}=U^\mu\eta_\mu{}^{\alpha\beta\gamma}$ and $P(U)_{\mu}{}^{\nu}\nabla_\nu=\nabla(U)_\mu$ the spatial covariant derivative \footnote{This definition should be modified when the derivative is acting on non-spatial fields.}.

Fourth, 
\beq
z(U)\cdot Z(U)=-a(U)\cdot E(U)=\frac{2M^2}{r^5 f^{5/2}}\,,
\eeq
where the dot here means left contraction of the vector and the 2-tensor.

Eq. \eqref{div_weyl} above in its split version (see Eqs. (2.5) in Ref. \cite{Bini:2004qf}, modulo a sign in $\rho^{(G)}$ depending on the sign convention used for Ricci) then becomes
\bea
f^{3/2}[{\rm div}(U)(f^{-3/2}E(U))]&=& \frac{4M^2}{r^5 f^{5/2}}  e_{\hat r}\,, \nonumber\\
f^{-1/2}{\rm Scurl}(U)[f^{1/2}E(U)]&=& 0\,,\nonumber\\ 
\eea
and can be further manipulated by using the identities
\bea
{\rm Scurl}(U)[k X]^{ab}&=& [\nabla(U) k \times X]^{ab} +k [{\rm Scurl}(U)[X]]^{ab}\nonumber\\
{}[{\rm div}(U)(k X)]^a &=& \nabla(U)_\mu k X^{\mu a}+k [{\rm div}(U)(X)]^a\,,
\eea
where $k$ is an arbitrary function one finds
\bea
 {\rm div}(U)(f^{-1/2}E(U)) &=& 0\,,\nonumber\\
{\rm Scurl}(U)[E(U)]&=& 0\,.
\eea
These Maxwell-like equations (Scurl-free electric field and divergence-free a rescaled version of it) furher characterize the geometry of the neutral $W$-soliton, with the additional Kerr-like property that the Simon tensor vanishes identically.

\section{Data availability} 
The data that support the findings of this article are openly available \cite{dataval}.

\end{document}